\documentclass[aps,prl,twocolumn,superscriptaddress,notitlepage]{revtex4-1}
\usepackage{graphicx}
\usepackage[caption=false]{subfig}
\usepackage{float}
\usepackage{natbib}
\usepackage{xcolor}
\usepackage{xr}
\usepackage{amsmath}
\usepackage{amssymb}
\usepackage{array}
\usepackage{wasysym}
\usepackage{color,soul}
\usepackage{braket}
\usepackage{verbatim}
\usepackage{hyperref}
\usepackage{gensymb}
\usepackage{url}
\usepackage{dirtytalk}

\usepackage{filecontents}
\begin{document}

\title{Correlated charge order intertwined with time-reversal symmetry-breaking nodal superconductivity in the dual flat band kagome superconductor CeRu$_{3}$Si$_{2}$}

\author{O. Gerguri}
\thanks{These authors contributed equally to the experiments.}
\affiliation{PSI Center for Neutron and Muon Sciences, 5232 Villigen PSI, Switzerland}
\affiliation{Physik-Institut, Universit\"{a}t Z\"{u}rich, Winterthurerstrasse 190, CH-8057 Z\"{u}rich, Switzerland}

\author{P. Kral}
\thanks{These authors contributed equally to the experiments.}
\affiliation{PSI Center for Neutron and Muon Sciences, 5232 Villigen PSI, Switzerland}

\author{M. Spitaler}
\affiliation{PSI Center for Neutron and Muon Sciences, 5232 Villigen PSI, Switzerland}

\author{M.~Salamin}
\affiliation{Department of Quantum Matter Physics, University of Geneva, CH-1211 Geneva, Switzerland}

\author{J.N. Graham}
\affiliation{PSI Center for Neutron and Muon Sciences, 5232 Villigen PSI, Switzerland}

\author{A. Doll}
\affiliation{PSI Center for Neutron and Muon Sciences, 5232 Villigen PSI, Switzerland}

\author{I.~Bia\l{}o}
\affiliation{Physik-Institut, Universit\"{a}t Z\"{u}rich, Winterthurerstrasse 190, CH-8057 Z\"{u}rich, Switzerland}

\author{I. Plokhikh}
\affiliation{TU Dortmund University, Department of Physics, Dortmund, 44227, Germany}
\affiliation{Research Center Future Energy Materials and Systems (RC FEMS), Germany}

\author{J. Krieger}
\affiliation{PSI Center for Neutron and Muon Sciences, 5232 Villigen PSI, Switzerland}

\author{T. J. Hicken}
\affiliation{PSI Center for Neutron and Muon Sciences, 5232 Villigen PSI, Switzerland}

\author{J. Oppliger}
\affiliation{Physik-Institut, Universit\"{a}t Z\"{u}rich, Winterthurerstrasse 190, CH-8057 Z\"{u}rich, Switzerland}

\author{L. Martinelli}
\affiliation{Physik-Institut, Universit\"{a}t Z\"{u}rich, Winterthurerstrasse 190, CH-8057 Z\"{u}rich, Switzerland}

\author{A. Steppke}
\affiliation{PSI Center for Neutron and Muon Sciences, 5232 Villigen PSI, Switzerland}

\author{N. Shepelin}
\affiliation{PSI Center for Neutron and Muon Sciences, 5232 Villigen PSI, Switzerland}

\author{R. Khasanov}
\affiliation{PSI Center for Neutron and Muon Sciences, 5232 Villigen PSI, Switzerland}

\author{M.v. Zimmermann}
\affiliation{Deutsches Elektronen-Synchrotron DESY, Hamburg, Germany}

\author{B. Monserrat}
\affiliation{Department of Materials Science and Metallurgy, University of Cambridge, Cambridge, UK}

\author{H. Luetkens}
\affiliation{PSI Center for Neutron and Muon Sciences, 5232 Villigen PSI, Switzerland}

\author{J.~Chang}
\affiliation{Physik-Institut, Universit\"{a}t Z\"{u}rich, Winterthurerstrasse 190, CH-8057 Z\"{u}rich, Switzerland}

\author{F.O.~von~Rohr}
\affiliation{Department of Quantum Matter Physics, University of Geneva, CH-1211 Geneva, Switzerland}

\author{Sun-Woo Kim}
\email{sunwookim@hanyang.ac.kr}
\affiliation{Department of Materials Science and Metallurgy, University of Cambridge, Cambridge, UK}
\affiliation{Department of Physics, Hanyang University, Seoul 04763, Republic of Korea}

\author{Z. Guguchia}
\email{zurab.guguchia@psi.ch}
\affiliation{PSI Center for Neutron and Muon Sciences, 5232 Villigen PSI, Switzerland}

\date{\today}

\begin{abstract}

\textbf{Kagome materials provide a powerful platform for exploring how flat electronic bands promote symmetry-breaking quantum states, yet studies have so far focused mainly on kagome-derived $d$-electron flat bands. In this paper, we introduce CeRu$_{3}$Si$_{2}$, a kagome superconductor in which our first-principles calculations show the coexistence of Ru $d$-orbital kagome flat bands and heavy-fermion flat bands derived from Ce$^{4+}$ $4f$-states. X-ray diffraction reveals a dominant 1/2 charge order with a much weaker 1/3 component persisting up to room temperature. Theoretical calculations further highlight the correlated nature of these charge-order states. Deep within the charge-ordered state, magnetoresistance emerges below 80 K and strengthens further below 30 K. Zero-field muon spin-rotation measurements show no time-reversal symmetry (TRS) breaking in the normal state, in contrast to LaRu$_{3}$Si$_{2}$ and YRu$_{3}$Si$_{2}$. However, an applied magnetic field induces weak magnetism. Across the $A$Ru$_{3}$Si$_{2}$ family ($A$ = La, Y, and Ce), the superconducting transition temperature $T_{\rm c}$ scales linearly with the onset temperature of normal-state TRS breaking $T_{\rm {TRSB}}$ and the magnitude of the field-induced magnetic response, revealing a direct positive correlation between normal-state symmetry breaking and superconductivity. Furthermore, we identify that CeRu$_{3}$Si$_{2}$ is the first 132-type kagome compound to host nodal superconductivity together with spontaneous internal magnetic fields, providing clear evidence for intrinsic TRS breaking in the superconducting state. These results establish CeRu$_{3}$Si$_{2}$ as a unique platform where intertwined kagome $d$- and heavy fermion $f$-electron flat bands generate a rich hierarchy of electronic orders. More broadly, they introduce a new paradigm for engineering correlated states through the interaction of distinct flat-band systems, with implications extending beyond kagome superconductors.}

\end{abstract}
\maketitle

\begin{figure*}[!]
    \centering
    \includegraphics[width=1\linewidth]{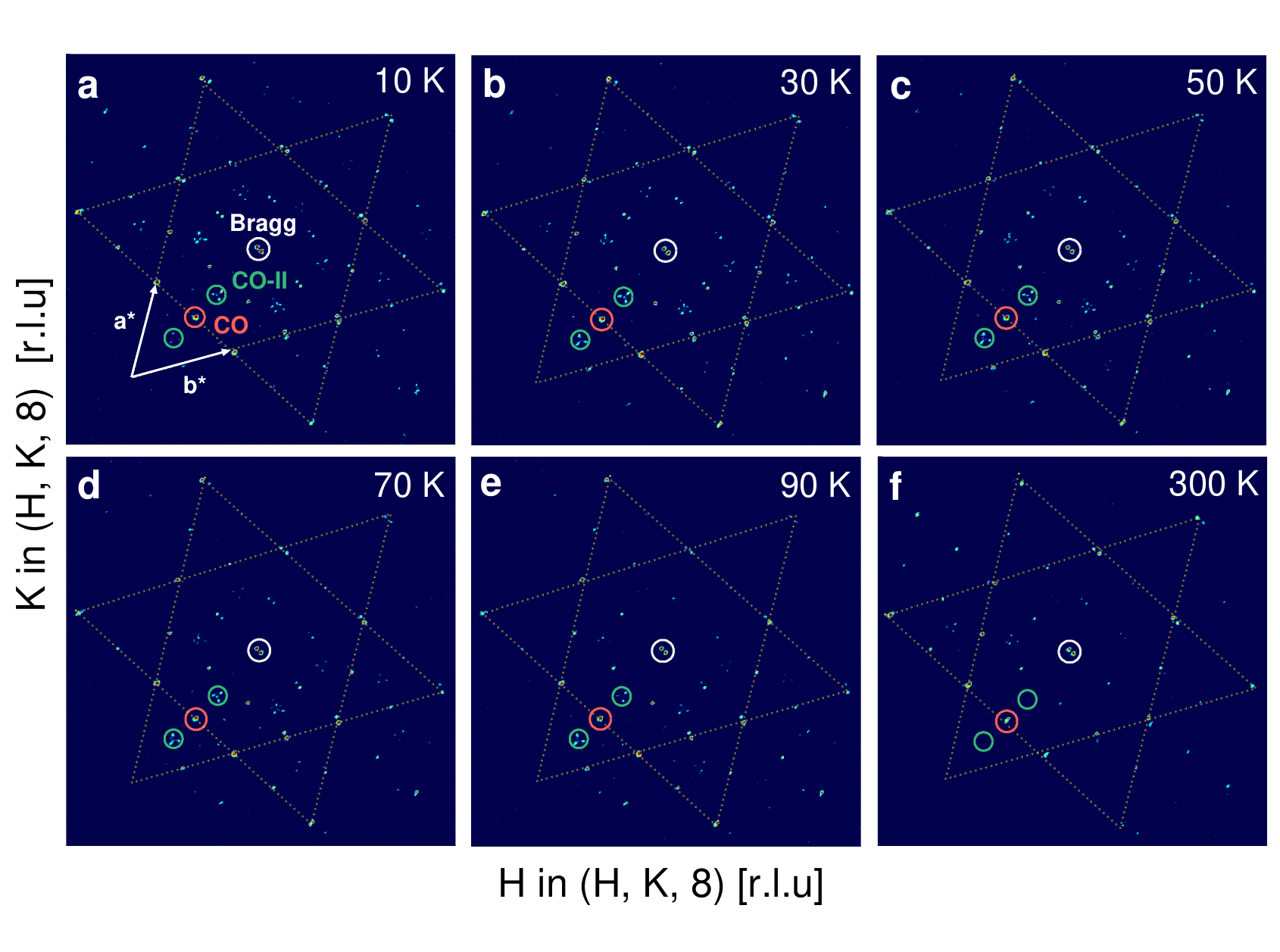}
    \caption{\textbf{Synchrotron X-ray diffraction experiments on a single crystal of CeRu$_{3}$Si$_{2}$.} (a-f) Reconstructed reciprocal space along the (0 0 1) direction at 8 r.l.u. (reciprocal lattice units), obtained at various temperatures between 10 K and 300 K. The main reflections are split, due to the formation of orthorhombic domains. Arrows indicate the reciprocal lattice vectors. White, red and green circles mark the Bragg peak, primary charge order CO peak, and the secondary charge order CO-II peak, respectively.}
    \label{fig:xrd}
\end{figure*}

\begin{figure*}[!]
    \centering
    \includegraphics[width=1\linewidth]{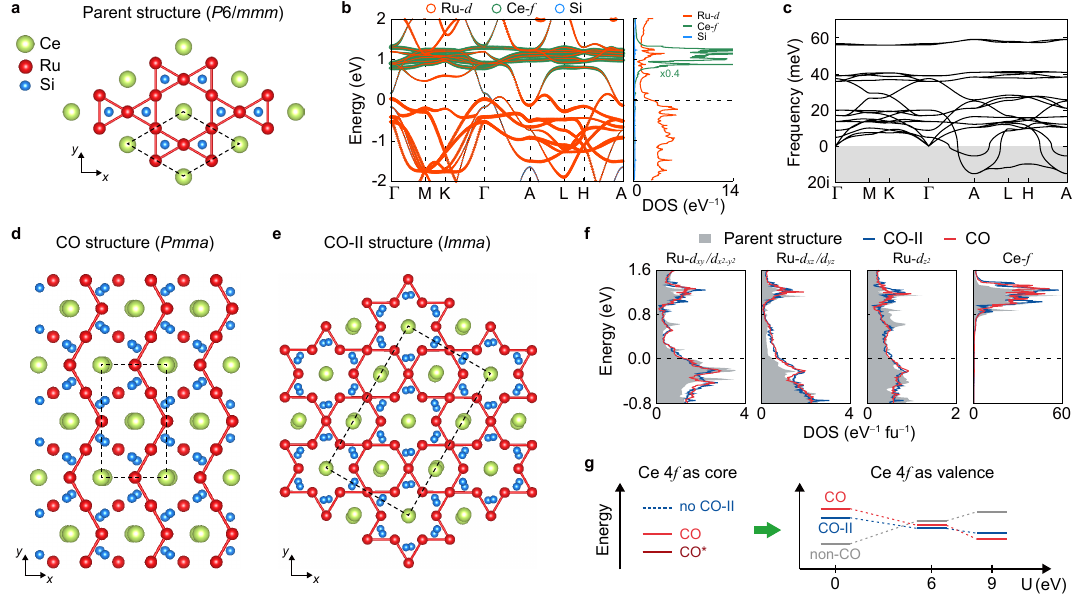}
    \caption{\textbf{Correlated charge-ordered states in CeRu$_{3}$Si$_{2}$.} (a) Atomic structure of the parent phase ($P6/mmm$). The dashed line indicates the primitive unit cell. (b) Orbital-projected band structure and density of states (DOS) of the parent structure. The sizes of the differently colored open circles overlaid on the band structure are proportional to the orbital-projected weights of the corresponding band states. (c) Phonon dispersion of the parent structure. (d,e) Atomic structures of the CO ($Pmma$) (d) and CO-II ($Imma$) (e) phases. (f) Orbital-projected DOS for the three structures. (g) Schematic energy diagram of the CO states obtained by treating the Ce $4f$ electrons as core (left) and valence (right). In the left panel, CO-II is not stabilized (relaxing into a different structure), and CO* denotes a charge order with $q=1/4$. In the right panel, the energies are plotted as a function of the Hubbard $U$ on the Ce $4f$ orbitals. Non-CO refers to a distorted structure with negligible Ru distortions.
    }
    \label{fig:theory}
\end{figure*}

\section{Main}

The interplay between crystal structure and electron correlations gives rise to a wide range of emergent electronic phases. Kagome lattices are a prime platform, hosting van Hove singularities \cite{kiesel2012sublattice,Kang_2022_twofold,Kim2023_monolayer,lin2024uniaxial}, flat bands \cite{wang2023quantum,mielke2021nodeless,guguchia2023tunable,jovanovic2022simple,kang2020dirac,gerguri2025distinct}, Dirac fermions \cite{ye2018massive,kang2020dirac}, pair density waves \cite{chen2021roton}, unconventional superconductivity \cite{guguchia2023unconventional,guguchia2023tunable}, and time-reversal symmetry breaking (TRSB) \cite{guguchia2023unconventional,guguchia2023tunable,mielke2022time,jiang2021unconventional,mielke2025coexisting,huang2026magnetic}.
A prominent example is the ternary kagome family $A$Ru$_{3}$Si$_{2}$ ($A$ = La, Y). LaRu$_{3}$Si$_{2}$ \cite{plokhikh2024discovery,mielke2021nodeless,ma2025correlation,li2025superconducting,deng2025theory,kral2026uniaxial} exhibits a dome-shaped superconducting phase diagram with one of the highest $T_{\rm c}$ values among kagome superconductors under low pressure, alongside multiple charge orders, field-induced magnetism below 80 K, and TRSB \cite{mielke2025coexisting} below 35 K. Substituting Y for La introduces chemical pressure, lowering $T_{\rm c}$ to $\sim$3.4 K while strengthening charge order; YRu$_{3}$Si$_{2}$ \cite{kral2026chargeorder} also shows TRSB below 25 K and field-induced magnetism below 90 K.
Despite differing energy scales, both compounds share nodeless superconductivity, high-temperature charge order, magnetoresistance, and normal-state TRSB. These phenomena arise from a common electronic structure with kagome-derived flat bands of Ru $d$-electrons, underscoring the central role of flat-band-driven correlations in shaping their phase diagrams.

In the above-mentioned and other kagome materials, attention has so far focused almost exclusively on $d$-electron flat bands. A fundamental open question is whether qualitatively new physics can arise when flat bands of distinct origin—such as kagome 
$d$-electron and heavy-fermion $f$-electron states—interact within a single system. To address this, we investigate the role of incorporating  $f$-electron degrees of freedom by focusing on CeRu$_{3}$Si$_{2}$, in which the superconducting transition temperature is further reduced to $T_{\rm c}$ ${\sim}$ 1 K. Previous investigations \cite{li2016chemical,rauchschwalbe1985investigation,rauchschwalbe1984superconductivity,yomo1986high,gulacsi1994bcs,kishimoto2003mixed} of CeRu$_{3}$Si$_{2}$ have been largely restricted to bulk measurements, such as magnetic susceptibility and specific heat. As a result, the electronic structure, the contribution of Ce $4f$ electrons to the density of states, the superconducting gap symmetry, and the possible emergence of charge order or anomalous normal-state electronic responses have remained under-explored. Here, our central goal is to elucidate how the strong suppression of $T_{\rm c}$ correlates with changes in the normal-state properties, to determine whether Ce $4f$ electrons contribute magnetically or via electronic correlations related to flat-bands. This would help to establish whether a fundamental link exists between superconductivity and normal-state electronic responses across the kagome family \textit{A}Ru$_{3}$Si$_{2}$ (\textit{A} = La, Y, and Ce). In this work, we present a comprehensive study of CeRu$_{3}$Si$_{2}$ over a wide temperature range, combining X-ray diffraction (XRD), magnetotransport, ${\mu}$SR, and first-principles calculations to uncover the rich landscape of electronic phases and to elucidate the influence of both heavy-fermion and kagome flat bands on the superconducting and normal-state properties.

\section{Results and discussion}
\subsection{Correlated Charge Orders Revealed by X-ray Diffraction and DFT Calculations}

Firstly, to determine the presence of charge order, a synchrotron \cite{ivashko2025p21} X-ray diffraction (XRD) experiments on CeRu$_{3}$Si$_{2}$ were conducted over a temperature range of 10 K to 300 K. The reconstructed reciprocal patterns along (0 0 1) for L = 8 r.l.u, obtained at various temperatures are shown in Fig.\ref{fig:xrd}a-f. From 300 K down to 10 K, 1/2 ($q_{1}$ = (1/2, 0, 0), $q_{2}$ = (1/2, 1/2, 0) and $q_{3}$ = (1/2, -1/2, 0)) and 1/3 ($q_{1}$ = (1/3, 0, 0), $q_{2}$ = (1/3, 1/3, 0) and $q_{3}$ = (1/3, -1/3, 0)) superstructure reflections emerge. The reciprocal vectors are given relative to the a${\times}$a${\times}$2c unit cell, according to the notation in Ref. \cite{plokhikh2024discovery}. This establishes the presence of various charge ordering within the orthorhombic phase in CeRu$_{3}$Si$_{2}$ at least below 300 K. Although in the current work, we did not explore the temperature region above 300 K, the direct siblings of CeRu$_{3}$Si$_{2}$, the $R$Ru$_{3}$Si$_{2}$ phases, undergo a transformation from a high-temperature $P6/mmm$ to low-temperature orthorhombic phases at temperature as high as 600 K for $R$ = La\cite{plokhikh2024discovery,mielke2025coexisting}, 760 K for $R$ = Nd \cite{misawa2025successive} and over 800 K for $R$ = Y \cite{kral2026chargeorder}. Hence, the observation of all three symmetry-related arms of charge-order vectors as well as pseudo-hexagonal symmetry of the diffraction pattern are due the formation of three symmetry related orthorhombic domains oriented according to $C_{3}$ symmetry of the high-temperature $P6/mmm$ phase ($a$${\times}$$a$${\times}$$c$) rather then 3$q$ order. Similar to YRu$_{3}$Si$_{2}$, the 1/2 order can be described within the $Pmma$ space group, while the 1/3 order, as in NdRu$_{3}$Si$_{2}$, corresponds to the $Imma$ space group. The instability of the $P6/mmm$ phase and its tendency to transform into low-temperature, lower-symmetry (orthorhombic) structures is further supported by DFT calculations.

To gain deeper insight into the charge-ordered states in CeRu$_{3}$Si$_{2}$, we performed first-principles DFT calculations (Fig. \ref{fig:theory}). These calculations reveal several key features of the electronic structure. In particular, they show the coexistence of Ru $d$-orbital kagome flat bands and heavy-fermion–like flat bands originating from the Ce$^{4+}$ $4f$ states, with the Fermi level located between these two sets of bands (Fig. \ref{fig:theory}b). The $f$ orbitals form flat bands situated above the Fermi level, consistent with a $4f^{0}$ configuration. The corresponding phonon dispersion shows multiple imaginary modes in the $q_z=1/2$ plane (Fig. \ref{fig:theory}c), which leads to two competing charge-order patterns: a $q$=1/2 $Pmma$ charge order (CO) and a $q$=1/3 charge order with $Imma$ symmetry (CO-II), as displayed in Figs. \ref{fig:theory}d,e. The CO structure exhibits a stripe-like pattern of in-plane Ru atoms while the CO-II structure shows a corner-sharing Star-of-David pattern. These distortions are accompanied by a suppression of density of states near the Fermi level associated with the in-plane Ru $d_{xy}$ and $d_{x^2-y^2}$ orbitals that comprise the Ru-$d$ flat bands (Fig. \ref{fig:theory}f). We also find that the Ce-$f$ flat bands shift upward upon the formation of the charge ordered states, indicating that both the Ru-$d$ and Ce-$f$ flat bands are involved in the charge-ordering instability.

\begin{figure*}[!]
    \centering
    \includegraphics[width= 1\linewidth]{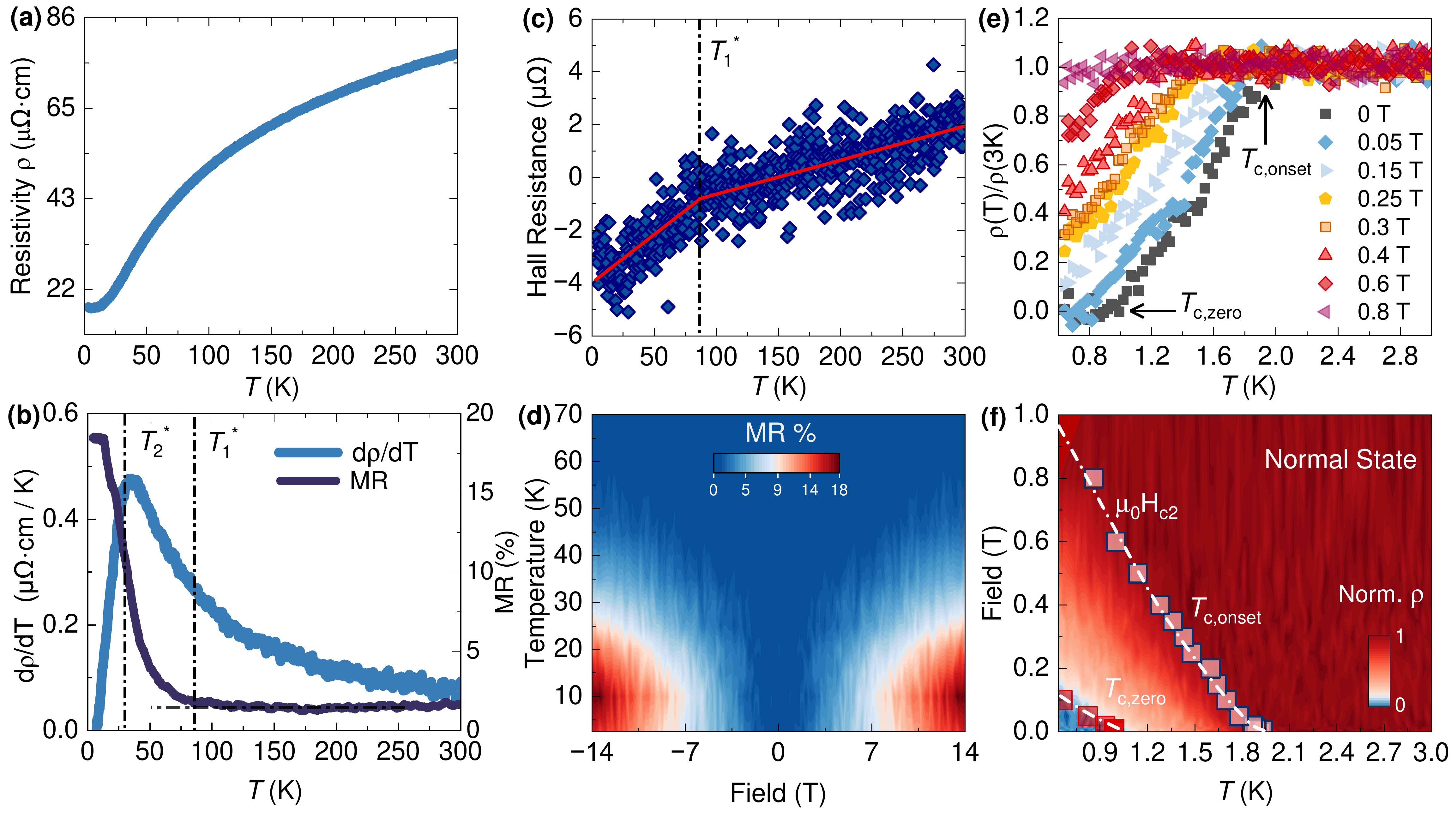}
    \caption{\textbf{Magnetotransport experiments on CeRu$_{3}$Si$_{2}$.} (a-c) Temperature dependence of the normal state resistivity $\rho$ (a), its first derivative (d$\rho$/d\textit{T}, left axis) and magnetoresistance (MR, right axis) (b), and Hall resistance measured at 14~T (c). Dashed vertical lines mark two characteristic temperatures, $T_{1}^{*}$ and $T_{2}^{*}$. $T_{1}^{*}$ corresponds to the onset of magnetoresistance and the sign reversal of the Hall resistance, while $T_{2}^{*}$ marks the peak in d$\rho$/d\textit{T}.
    (d) Color plot of the normal state MR, drawn  as a function of temperature and applied field (up to 14~T). (e) Temperature dependence of the resistivity across the superconducting transition, measured down to 0.6 K under magnetic fields up to 0.8 T. Arrows indicate temperatures corresponding to the onset of the superconducting transition and the zero-resistance state. (f) Field–temperature color map of the resistivity derived from the data in (e). Pink squares indicate the estimated onset of the superconducting transition, $\textit{T}_{\rm{c},\rm{onset}}$, while red squares mark the zero-resistance superconducting state at $\textit{T}_{\rm{c},\rm{zero}}$. Dashed lines are guides to the eye.}
    \label{fig:transport}
\end{figure*}

\begin{figure*}[!]
    \centering \includegraphics[width = 0.9 \linewidth]{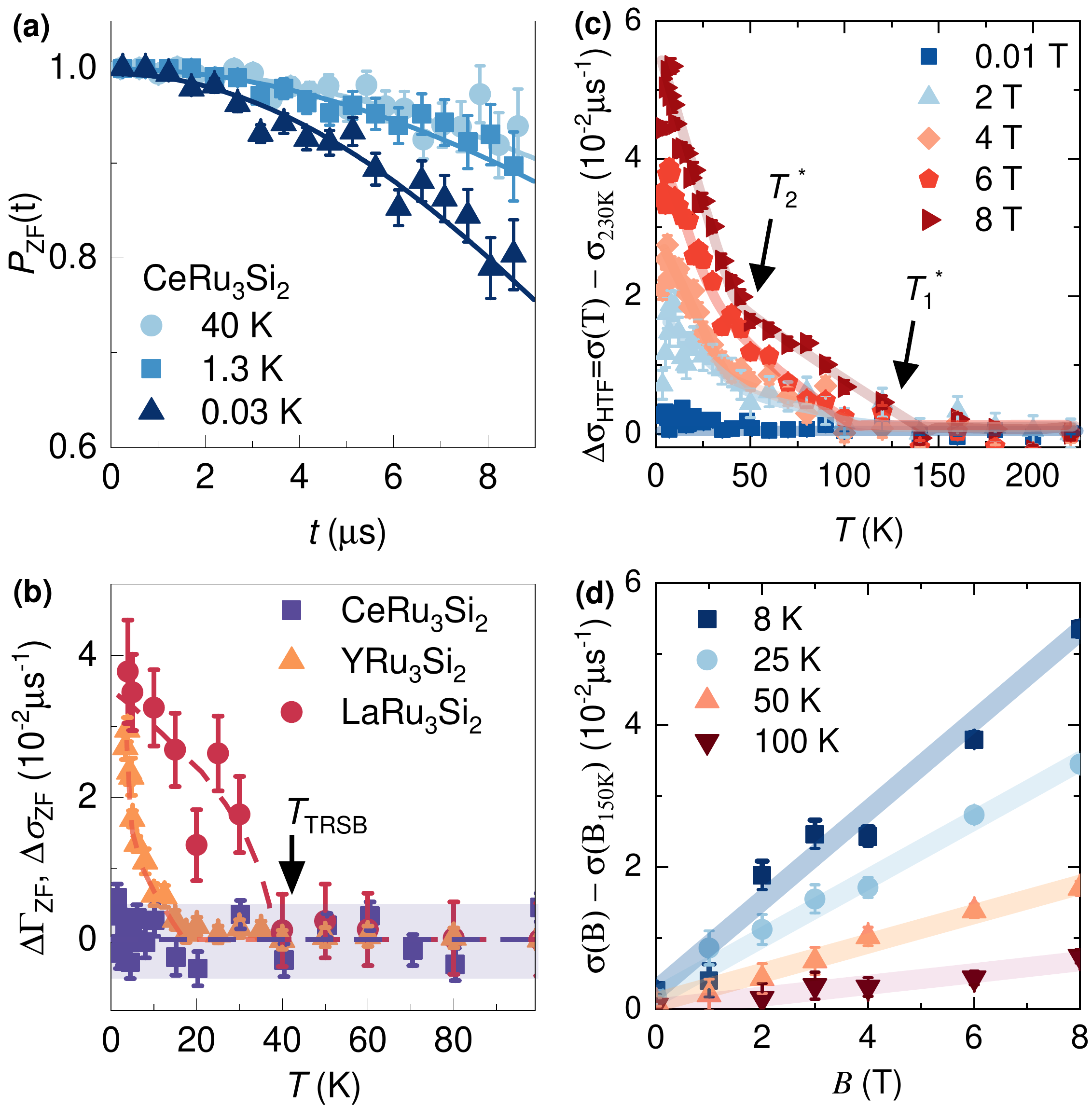}
    \caption{\textbf{Zero- and high-field $\mu$SR experiments on CeRu$_{3}$Si$_{2}$.} 
    (a) The ZF-$\mu$SR time spectra obtained at three different temperatures, $T$= 0.03 K, 1.3 K, and 40 K, covering the temperature range both below and above $T_{\rm c}$. The solid curves represent fits to the recorded time spectra, using the Gaussian Kubo Toyabe (GKT) function. (b) Temperature dependence of the zero-field $\mu$SR relaxation rates, $\Delta$$\Gamma$$_{\rm{ZF}}$ and $\Delta$$\sigma$$_{\rm{ZF}}$, shown after subtraction of the high-temperature contribution. Because the zero-field ${\mu}$SR spectra of the Y and La samples exhibit an exponential form, with the corresponding relaxation rate $\Gamma$ increasing at low temperatures, we present $\Delta$$\Gamma$$_{\rm{ZF}}$. In contrast, the Ce sample shows Gaussian depolarization in zero field, and thus the Gaussian relaxation rate $\Delta$$\sigma$$_{\rm{ZF}}$ is presented. (c)  Temperature dependence of the transverse-field $\mu$SR relaxation rate $\Delta$$\sigma_{\rm{HTF}}$, recorded under various applied magnetic fields up to 8 T, shown after subtraction of the high-temperature contribution. Arrows mark two characteristic temperatures $T_{1}^{*}$ and $T_{2}^{*}$. (d) Field dependence of $\Delta$$\sigma_{\rm{HTF}}$ at four selected temperatures below $T_{1}^{*}$.}
    \label{fig:HF}
\end{figure*}

Further analysis reveals that the Ce-$4f$ states play a crucial role in determining the charge-ordered ground state of this compound (Fig. \ref{fig:theory}g), despite the $f$ orbitals forming flat bands above the Fermi level. While the Hubbard interaction $U$ on the Ce $4f$ orbitals has only a minor influence on the lattice parameters (less than 0.4\% change from $U=0$ to 9\,eV), it strongly affects the relative energetics of the charge-ordered states, demonstrating the essential role of electronic correlations in stabilizing the charge order. Specifically, when the Ce $4f$ states are excluded from the valence (i.e., treated as core electrons), the $q$=1/3 order is completely suppressed and the $q$=1/2 order is not stabilized as the lowest energy ground state, in contradiction with experimental observations. In contrast, when the Ce $4f$ electrons are included as valence states, the $q$=1/3 order is energetically favored over the $q$=1/2 order at $U$=0\,eV. However, the ground state is still a different structure with negligible Ru distortions, which we classify as non-CO order. The two charge-ordered states become nearly degenerate at $U$=6\,eV, and for $U$ ${\textgreater}$ 6\,eV the $q$=1/2 order becomes the lowest-energy ground state, while the $q$=1/3 order remains very close in energy. These results suggest that a relatively large Hubbard interaction ($U$ ${\textgreater}$ 6\,eV) for the Ce-$f$ states is required to reproduce the experimental observations, highlighting the correlated nature of the dominant $q$=1/2 charge order together with a weaker $q$=1/3 component observed experimentally.


\begin{figure*}[!]
    \centering
    \includegraphics[width=1\linewidth]{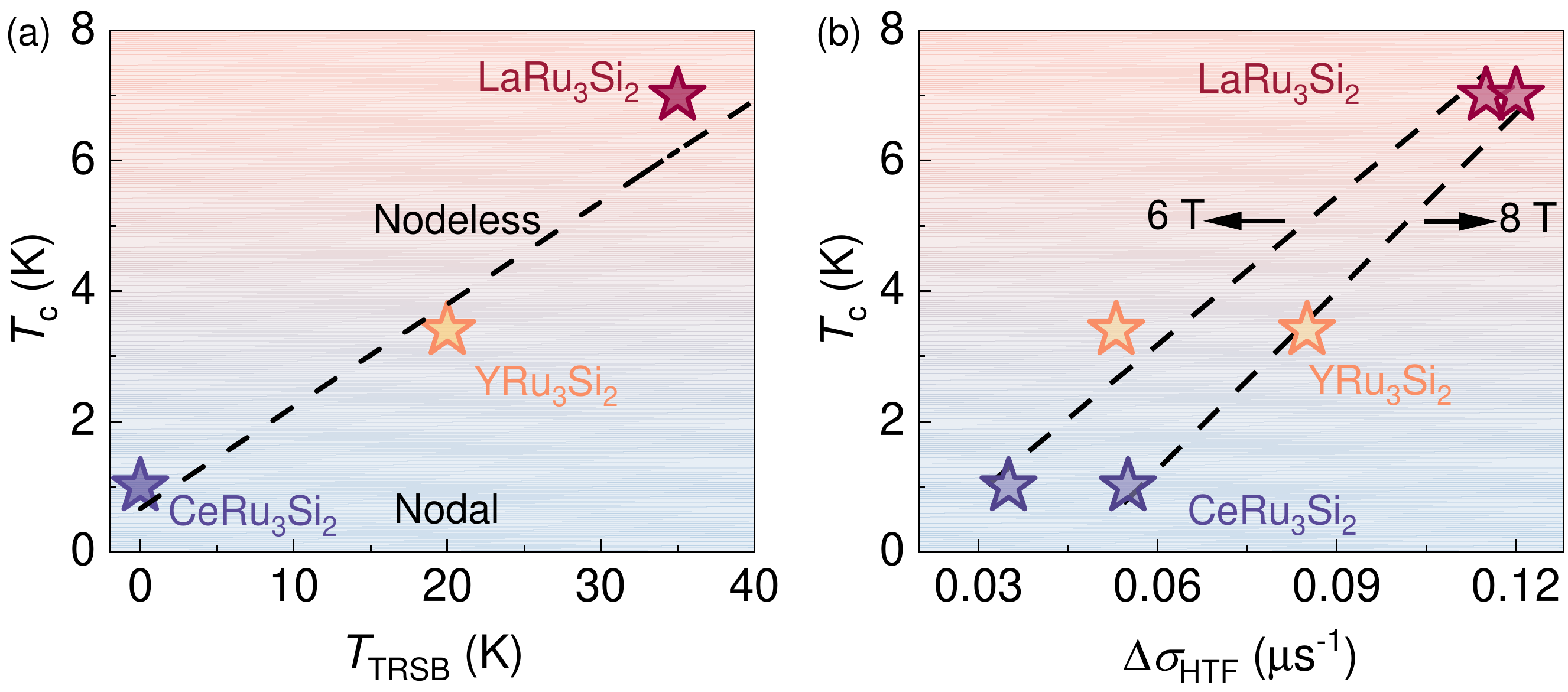}
    \caption{\textbf{Scaling between superconducting and normal state properties across the \textit{A}Ru$_{3}$Si$_{2}$ series (\textit{A} = Ce, Y, La).} (a,b) Superconducting transition temperature $T_{\rm c}$ plotted as a function of the zero-field normal-state TRS breaking onset temperature, $T_{\rm TRSB}$, (a) and the absolute value of the field-induced relaxation rate measured at 6 T and 8 T (b).}
    \label{fig:TRSB}
\end{figure*}

\subsection{Magnetoresistance, Hall Reversal, and Superconducting Properties}

Having established the presence of correlated charge order and supported it by DFT calculations, we now examine the transport properties in both the normal and superconducting states. Figure~\ref{fig:transport}a shows the temperature dependence of the electrical resistivity of CeRu$_{3}$Si$_{2}$ above $T_{ \rm{c}}$. In the charge-ordered phase, the resistivity remains metallic, consistent with behavior reported for LaRu$_{3}$Si$_{2}$\cite{mielke2025coexisting}, YRu$_{3}$Si$_{2}$\cite{kral2026chargeorder}, and the kagome superconductors AV$_{3}$Sb$_{5}$\cite{liu2025,xu2025av3sb5}. This indicates that charge ordering in CeRu$_{3}$Si$_{2}$ only induces a partial reconstruction of the Fermi surface, leaving a substantial fraction of itinerant carriers intact. Electrical transport is dominated by ungapped bands, resulting in metallic behavior upon cooling. This is typical of multiband or three-dimensional systems\cite{mansart2012evidence} and contrasts with quasi-one-dimensional Peierls systems, where charge order\cite{ong1977anomalous} opens a large gap and suppresses conductivity. The resistivity curve ${\rho}(T)$ (Fig. \ref{fig:transport}a) reveals two characteristic temperature scales governing the normal-state electronic properties. The first derivative of ${\rho}(T)$ (Fig.~\ref{fig:transport}b) exhibits a weak anomaly at $T_{1}^{*}$ ${\simeq}$ 90 K, followed by a pronounced maximum at  $T_{2}^{*}$ ${\simeq}$ 35 K. Motivated by these features, we carried out systematic measurements of the magnetoresistance (MR) (Fig.~\ref{fig:transport}b and \ref{fig:transport}d) and Hall resistivity (Fig.~\ref{fig:transport}c) over a broad range of temperatures and magnetic fields. The  MR at 14 T becomes finite below ${\simeq}$ 90~K and reaches values up to ${\simeq}$ 18${\%}$, which is consistent with the positive MR expected in metallic systems\cite{ali2014large}. Two clear slope changes can be discerned: an initial onset below $T_{1}^{*}$ and a much steeper increase below $T_{2}^{*}$, closely tracking the anomalies in the resistivity derivative. Concurrently, the Hall resistivity undergoes a sign change across $T_{1}^{*}$. Taken together, these magnetotransport signatures point to a normal-state transition or crossover occurring between the onset of charge order and the emergence of superconductivity. Figure~\ref{fig:transport}e shows the superconducting transitions under applied magnetic fields. For completeness, the data are also presented as a color map (Fig.~\ref{fig:transport}f) with temperature on the horizontal axis and magnetic field on the vertical axis, where blue denotes the zero-resistance superconducting state and red the normal state. The superconducting state is expected to be progressively suppressed by a magnetic field as noted by the shift to lower transition temperatures (see Fig.~\ref{fig:SC}e,f), yielding an estimated upper critical field H$_{\rm{c2}}$ of 1~T at 0.6~K. The transition exhibits an onset at 2~K, while zero resistance is achieved only below 0.9~K, resulting in an unusually broad temperature interval without zero resistance. X-ray diffraction confirms a homogeneous single phase, excluding sample inhomogeneity as the origin of this behavior. Moreover, as shown below in Fig.~\ref{fig:SC}d, ${\mu}$SR detects the onset of superconductivity only below 0.9~K. Given its high sensitivity even to small superconducting volume fractions, this rules out a distribution of transition temperatures, which would otherwise produce a higher-temperature onset. Instead, the higher-temperature onset of the transition likely reflects two-dimensional superconductivity\cite{qiu2021recent}, with full three-dimensional phase coherence established only below 0.9~K. This scenario warrants further investigation in single crystals, once available.

\subsection{Nodal TRS-Breaking Superconductivity and SC–Normal-State Universal Scaling}

Next, we present ${\mu}$SR results probing the normal state of CeRu$_{3}$Si$_{2}$. A central question is whether magnetism is associated with the characteristic temperature scales $T_{1}^{*}$ and $T_{2}^{*}$ identified in transport, as previously observed in LaRu$_{3}$Si$_{2}$ and YRu$_{3}$Si$_{2}$. To address this issue, we combine zero-field (ZF) and high transverse-field (HTF) ${\mu}$SR measurements. The ZF-${\mu}$SR spectra display only a very weak depolarization of the muon spin ensemble (Fig.~\ref{fig:HF}a), providing no evidence for long-range magnetic order in CeRu$_{3}$Si$_{2}$. Full muon polarization is recovered upon application of a small longitudinal field ($B_{{\rm LF}}$~=~5~mT), indicating that the relaxation originates from weak, internal fields quasi-static on the microsecond timescale. The data are well described by a Gaussian Kubo–Toyabe function, consistent with previous reports. Importantly, the relaxation rate remains temperature independent down to 1.3 K (Fig.~\ref{fig:HF}b), excluding time-reversal symmetry (TRS) breaking across both $T_{1}^{*}$ and $T_{2}^{*}$. This demonstrates that the normal state of CeRu$_{3}$Si$_{2}$ preserves TRS, in contrast to its La and Y counterparts (Fig.~\ref{fig:HF}b)\cite{mielke2025coexisting,kral2026chargeorder}.


To further corroborate these findings, we performed a comprehensive set of HTF-${\mu}$SR measurements\cite{sedlak2012musrsim} as seen in Fig.~\ref{fig:HF}c and d. At the lowest applied field of 0.01 T, $\sigma_{\rm HTF}$ remains nearly temperature independent down to 3 K, consistent with the ZF-${\mu}$SR results. With increasing magnetic field, however, a clear two-step enhancement of  $\sigma_{\rm HTF}$ emerges, with onsets at $T_{1}^{*}$ and a further increase below $T_{2}^{*}$, demonstrating a field-induced magnetic response that strengthens with field. Notably, the relaxation rate increases across the entire sample volume, which confirms the bulk nature of the magnetic response. These results establish that the magnetoresistance observed in CeRu$_{3}$Si$_{2}$ is associated with a weak, field-induced magnetic state. The absolute increase in relaxation rate, ${\Delta}$$\sigma_{\rm HTF}$ = $\sigma_{\rm HTF}(3\rm{K})$ - $\sigma_{\rm HTF}(\rm 150K)$, amounts to 0.053 ${\rm{\mu}}$$s^{-1}$, significantly smaller than in LaRu$_{3}$Si$_{2}$ and YRu$_{3}$Si$_{2}$. Thus, CeRu$_{3}$Si$_{2}$ is distinguished not only by its reduced superconducting transition temperature $T_{\rm c}$, but also by the absence of normal-state TRS breaking in zero field and by a substantially weaker field-induced magnetic response. To quantify this trend, we compare $T_{\rm c}$ with both the onset temperature of ZF-TRS breaking (Fig. \ref{fig:TRSB}a) and ${\Delta}$$\sigma_{\rm HTF}$ measured at 6 and 8 T (Fig. \ref{fig:TRSB}b) across the $A$Ru$_3$Si$_2$ series ($A$ = Ce, Y, La). Strikingly, $T_{\rm c}$ exhibits an approximately linear correlation with both quantities, indicating that superconductivity is closely linked to the strength of the normal-state magnetic response. Together, these results point to a common pairing mechanism and support an electronic origin of superconductivity in this kagome family.

\begin{figure*}[!]
    \centering
    \includegraphics[width=1\linewidth]{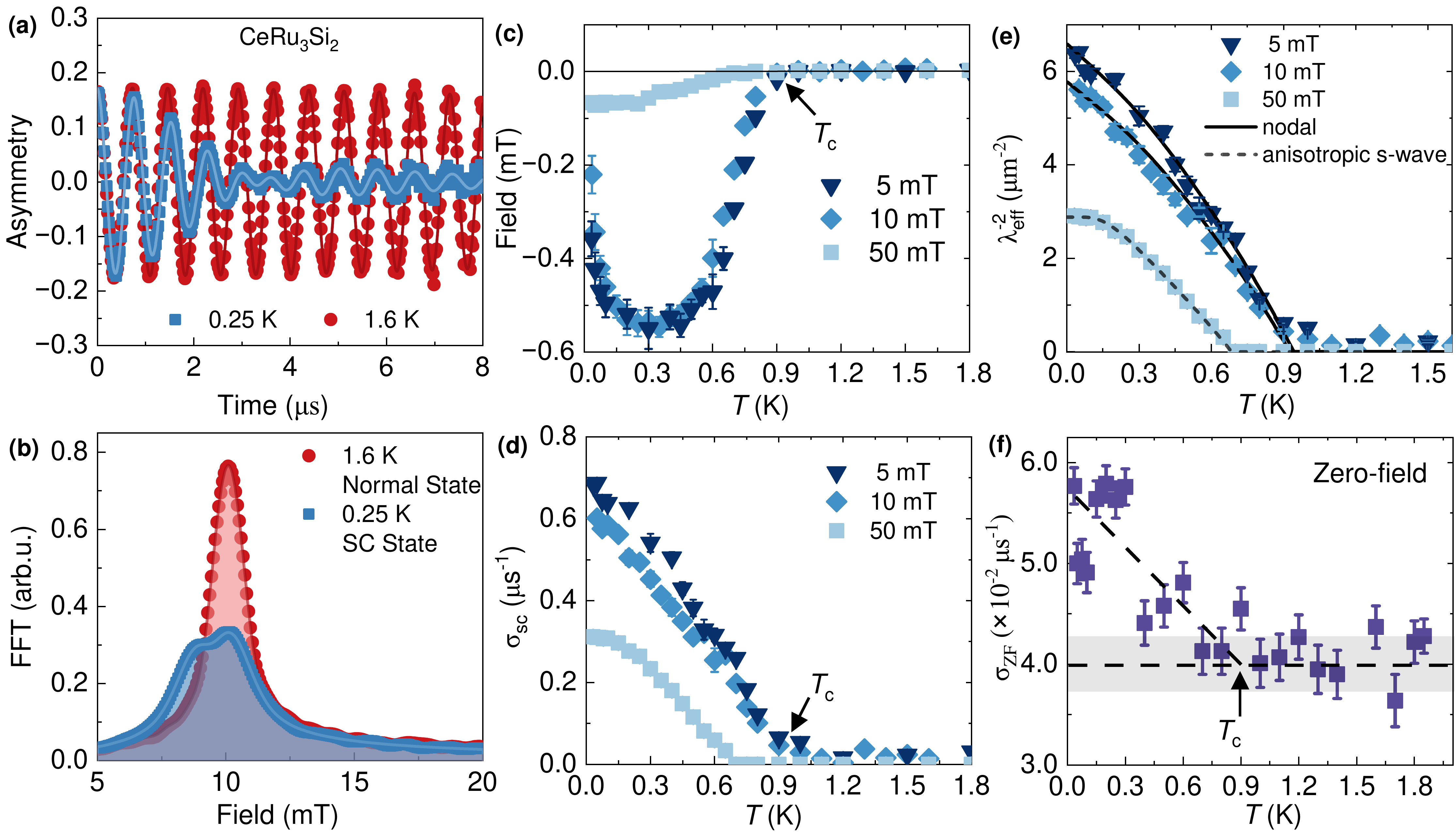}
    \caption{\textbf{$\mu$SR insights into the superconducting state of CeRu$_{3}$Si$_{2}$.} (a) Transverse-field (TF) $\mu$SR time spectra obtained above (1.6 K) and below (0.03 K) the SC transition under an applied magnetic field of 10 mT. The solid lines represent fits to the data using Eq.~1 (see Methods section). Error bars are the standard error of the mean (s.e.m.) in about 10$^{6}$ events. (b) Fourier transform of the measured time spectra in the normal and superconducting states. (c,d) Temperature dependence of the diamagnetic shift \(\Delta B_\text{int}(T)\) (c) and the muon spin depolarization rate \(\sigma_\text{\rm{sc}}(T)\) (d) under the applied magnetic fields of 5 mT, 10 mT, and 50 mT. The error bars represent the standard deviation of the fit parameters. (e) Inverse squared effective penetration depth $\lambda^{-2}_{\rm{eff}}$ as a function of temperature fitted with several theoretical models (see Methods section). (f) Temperature dependence of zero-field muon spin relaxation rate across $T_{\rm c}$. Dashed lines are guides to the eye. The shaded region indicates the range of normal-state relaxation rates $\sigma_{\rm{nm}}$, which remains constant up to 300 K.}
    \label{fig:SC}
\end{figure*}

In the following, we investigate the microscopic nature of the superconducting state in CeRu$_{3}$Si$_{2}$. Figure 6a shows representative transverse-field (TF) ${\mu}$SR time spectra recorded in an applied magnetic field of 10 mT above ($T$ = 1.6 K) and below ($T$ = 0.25 K) the superconducting transition. Above $T_{\rm c}$, the muon spin precession exhibits only weak damping, arising from the random local fields of nuclear moments. Upon entering the superconducting state, the damping rate increases markedly, reflecting the formation of a non-uniform internal magnetic field distribution associated with a flux-line lattice\cite{brandt1988flux}. This evolution is further illustrated in Fig. 6b, which displays the Fourier transforms of the ${\mu}$SR spectra at the corresponding temperatures. Above $T_{\rm c}$, a sharp and symmetric field distribution is observed. In contrast, in the superconducting state the distribution becomes significantly broadened and asymmetric, and shifts relative to the applied field--hallmarks of a vortex lattice in the mixed state. From these data, two key quantities are extracted: the diamagnetic shift \(\Delta B_\text{int}=\mu_0(H_\text{int, SC}-H_\text{int, NS})\), defined as the difference between the internal fields in the superconducting $H_{\rm{int},\rm{SC}}$ and normal states $H_{\rm{int}, \rm{NS}}$ and the total muon spin depolarization rate \(\sigma_\text{tot}\) (\(=\sqrt{\sigma_\text{SC}^2+\sigma_\text{nm}^2}\)), which includes contributions from the superconducting vortex lattice (\(\sigma_\text{SC}\)) and from nuclear dipolar fields (\(\sigma_\text{nm}\)). The temperature dependences of $\sigma_{\rm SC}$ and ${\Delta}$$B_{\rm int}$ for applied fields of 5, 10, and 50 mT are shown in Fig. 6c and 6d, respectively. The large diamagnetic response observed at all fields confirms the bulk nature of superconductivity. Notably, at low fields (5 and 10 mT) an additional upturn in ${\Delta}$$B_{\rm int}$ appears below ${\simeq}$ 0.3 K, which is completely suppressed at 50 mT. The disappearance of this feature with increasing field indicates that it is not of magnetic origin—since a magnetic contribution would be enhanced by field-but is instead associated with intrinsic properties of the vortex state or the superconducting gap structure. A corresponding field-dependent behavior is also evident in $\sigma_{\rm SC}$, which shows qualitatively different temperature dependences at low and high fields. To quantitatively analyze the superconducting gap structure, we calculated the temperature dependence of the superfluid density 1/${\lambda}^{2}$ (see Methods) and fitted the data using models describing a single isotropic full gap, an anisotropic nodeless gap, and a nodal gap. The results are presented in Fig. \ref{fig:SC}e. At low fields (5 and 10 mT), the data are well described by a nodal gap model, whereas at 50 mT the penetration depth clearly saturates at low temperatures and is consistent with an anisotropic nodeless gap. These results demonstrate a field-induced crossover from nodal to nodeless superconductivity in CeRu$_{3}$Si$_{2}$. Such nodal behavior at low magnetic fields is unique within the 132 kagome family, as LaRu$_{3}$Si$_{2}$ and YRu$_{3}$Si$_{2}$ exhibit well-defined nodeless gaps even at low fields. A further distinction between CeRu$_{3}$Si$_{2}$ and its La/Y counterparts is revealed by zero-field ${\mu}$SR measurements: below $T_{\rm c}$, the muon spin relaxation rate increases continuously by ${\simeq}$ 0.02 ${\mu}$$s^{-1}$ (Fig. 6f), corresponding to the appearance of spontaneous internal magnetic fields of order ${\simeq}$0.3 G. This provides clear evidence for time-reversal symmetry breaking in the superconducting state, while TRS remains preserved in the normal state. Taken together, the nodal superconducting gap at low fields, the field-induced crossover to a nodeless gap, the emergence of spontaneous internal fields below $T_{\rm c}$, and the low superfluid density establish CeRu$_{3}$Si$_{2}$ as an unconventional superconductor and a unique member of the 132 kagome family.\\

\section{Summary and Outlook}

Here, we summarize and provide deeper insight into each set of observations. By combining X-ray diffraction, magnetotransport, ${\mu}$SR, and first-principles calculations, we uncover a series of novel and closely intertwined properties of CeRu$_{3}$Si$_{2}$, establishing it as a highly unconventional kagome superconductor. Specifically, we present the following key findings: \textbf{(1)} First-principles calculations reveal the coexistence of Ru $d$-orbital kagome flat bands and heavy-fermion flat bands derived from Ce$^{4+}$ $4f$-states, highlighting an unusual interplay between kagome flat-band physics and strong electronic correlations. To our knowledge, CeRu$_{3}$Si$_{2}$ represents the first kagome superconductor in which heavy-fermion and kagome flat bands coexist and cooperate, providing a new route toward emergent correlated quantum states.
\textbf{(2)} X-ray diffraction uncovers a dominant 1/2 charge order accompanied by a much weaker 1/3 order, both persisting at least up to room temperature. First-principles calculations reproduce the observed charge-order patterns and indicate that a relatively large Hubbard interaction ($U$ ${\textgreater}$ 6 eV) for the Ce-$f$ states is required to account for the experimental observations. This finding highlights the strongly correlated nature of the dominant $q$=1/2 charge order, accompanied by a weaker $q$=1/3 component observed experimentally. While a 1/2 charge order has also been reported in YRu$_{3}$Si$_{2}$\cite{kral2026chargeorder} and NdRu$_{3}$Si$_{2}$\cite{misawa2025successive}, LaRu$_{3}$Si$_{2}$ instead hosts\cite{mielke2025coexisting} a 1/4 charge order below ${\simeq}$ 400 K and an additional 1/6 order below ${\simeq}$ 80 K. CeRu$_{3}$Si$_{2}$ thus constitutes the first kagome system exhibiting two distinct charge-ordering wave vectors already at room temperature. Taken together, these results demonstrate that the 132 kagome family provides a fertile platform for multiple competing and coexisting charge-ordered states.
\textbf{(3)} Deep within the charge-ordered phase, a finite magnetoresistance emerges below ${\simeq}$ 80 K and becomes strongly enhanced below ${\simeq}$ 30 K. This behavior is accompanied by a sign reversal of the Hall resistivity across ${\simeq}$ 80 K. Notably, no additional charge-order transitions are detected below this temperature, suggesting that the emergence of magnetoresistance and the Hall sign change arise from a reconstruction of the Fermi surface\cite{taillefer2009fermi} driven by electronic or magnetic correlations rather than by a new static charge order.
\textbf{(4)} Zero-field ${\mu}$SR measurements reveal no evidence for TRS breaking in the normal state, in sharp contrast to LaRu$_{3}$Si$_{2}$ and YRu$_{3}$Si$_{2}$. However, the application of a magnetic field induces a weak but bulk magnetic response, with an onset near ${\simeq}$~80~K and a pronounced enhancement below ${\simeq}$~30~K. The close correspondence between these temperature scales and the onset of magnetoresistance indicates that the magnetic and electronic responses are strongly intertwined in CeRu$_{3}$Si$_{2}$.
\textbf{(5)} Across the $A$Ru$_{3}$Si$_{2}$ series ($A$ = Ce, Y, La), the superconducting transition temperature $T_{\rm c}$ scales linearly with both the onset temperature of normal-state TRS breaking $T_{\rm TRSB}$ and the magnitude of the field-induced magnetic response. This striking correlation reveals a direct positive link between normal-state symmetry breaking, magnetic fluctuations, and superconductivity.
\textbf{(6)} CeRu$_{3}$Si$_{2}$ is identified as the first 132-type kagome compound to host nodal superconductivity at low magnetic fields, coexisting with spontaneous internal magnetic fields. This provides compelling evidence for intrinsic TRS breaking in the superconducting state and places CeRu$_{3}$Si$_{2}$ at the intersection of kagome flat-band physics, strong correlations, and unconventional superconductivity.
\textbf{(7)} We find an intriguing field dependence of the superconducting gap structure in CeRu$_{3}$Si$_{2}$, revealed by the temperature dependence of the superfluid density: at low magnetic fields (5 and 10 mT) the data are best described by a nodal gap, whereas at higher fields (50 mT) an anisotropic nodeless behavior emerges. Several scenarios may account for this crossover. A natural explanation is multiband superconductivity\cite{xu2021multiband,huang2020pairing}, in which a nodal or strongly anisotropic gap on one Fermi-surface sheet dominates the low-field response but is preferentially suppressed by magnetic field due to its smaller gap magnitude or lower upper critical field, leaving a fully gapped band to govern the superfluid response at higher fields. Alternatively, the nodes may be accidental rather than symmetry protected\cite{eremin2006magnetic}, in which case magnetic field can lift deep gap minima or nodes by modifying the relative weights of competing gap harmonics or by suppressing intertwined electronic orders. In this context, field-induced suppression of competing charge-ordered or pair-density-wave correlations—known to generate nodal-like quasiparticle spectra—could restore an uniform superconducting state with a full gap. Finally, field-dependent vortex-state effects and nonlinear quasiparticle contributions may also affect the apparent low-energy excitation spectrum. Together, these possibilities indicate that the observed nodal-to-nodeless evolution reflects a delicate interplay between multiband superconductivity, gap anisotropy, and competing electronic correlations in CeRu$_{3}$Si$_{2}$, and underscores the need for further field-resolved spectroscopic and thermodynamic studies to disentangle their relative roles.


Taken together, we establish CeRu$_{3}$Si$_{2}$ as a unique platform in which kagome-derived Ru $d$ flat bands and Ce $4f$ heavy-fermion flat bands coexist and act cooperatively, giving rise to a hierarchy of correlated phenomena—from robust correlated charge order up to room temperature, to large normal-state magnetoresistance, and nodal superconductivity with intrinsic time-reversal symmetry breaking. The observation of a universal linear scaling across the $A$Ru$_{3}$Si$_{2}$ ($A$ = Ce, Y, La) family, linking $T_{\rm c}$, the onset of normal-state TRS breaking, and the field-induced magnetic response, demonstrates that symmetry breaking in the normal state promotes superconductivity, pointing to an electronically driven pairing mechanism. More broadly, our results introduce a paradigm in which distinct flat-band systems coexist to generate intertwined quantum orders, opening a route to engineer novel correlated states by combining kagome and heavy-fermion physics within a single material platform.

\section{Methods}

\textbf{Sample Preparation:}
Polycrystalline samples of CeRu$_{3}$Si$_{2}$ were prepared by arc-melting
from mixtures of cerium chunks (99.9$\%$, Thermo Scientific), ruthenium powder (99.99$\%$, Leico), and silicon pieces (99.95$\%$, Sigma
Aldrich) under an argon atmosphere. Stoichiometric amounts of the
elements were used with a 30$\%$ molar excess of ruthenium to avoid
the formation of the {CeRu$_{2}$Si$_{2}$} phase. The ruthenium powder was
pressed into pellets to avoid sputtering, and the melting process was
started with the metals (cerium and ruthenium) so that the melt could absorb silicon. A piece of zirconium was used as a getter to remove residual oxygen. During the arc-melting process, the samples
were melted and flipped several times for better homogenization. The pellets were not shiny on the surface after the synthesis; this was due to the formation of a thin oxide layer, which was mechanically removed. All of the measurements were performed on CeRu$_{3}$Si$_{2}$
stemming from a single sample pellet.\\

\textbf{X-ray diffraction:}
We employed synchrotron hard X-ray diffraction (from 10 to 300\,K) to uncover the charge-ordered
state. This diffraction technique provides direct evidence of structural transitions and charge order. The
experiments are complemented by DFT calculations,
which provide microscopic insight into each state. Single
crystals of CeRu$_{3}$Si$_{2}$ smaller than 50 µm were obtained from the crushed arc-melted polycrystalline pellet, washed consequently in
dilute HCl, H$_{2}$O and ethanol and mounted on carbon fiber holders. 
Hard X-ray synchrotron experiments were carried out
at the P21.1 beamline\cite{ivashko2025p21} at Petra III, DESY, using
101.6 keV photons. These measurements were done with
the Pilatus 3 X CdTe 2M detector. To maintain temperature control down
to 10K and up to 300K, a Displex cryostat was used. Data collection, as well as reconstruction of reciprocal space, were performed using the beamlines P21.1  reconstruction algorithm.\\

\textbf{First-principles calculations:}
We performed density functional theory calculations using the Vienna ab initio simulation package {\sc vasp} \cite{VASP}, implementing the projector-augmented wave (PAW) method \cite{PAW}. We employed PAW pseudopotentials with the following valence configurations: for Ce, two different valence treatments are considered, $5s^2 5p^6 5d^1 6s^2$ (without $4f$ electrons) and $4f^1 5s^2 5p^6 5d^1 6s^2$ (including $4f$ electrons); for Ru atoms, $4s^2 4p^6 4d^7 5s^1$; and for Si atoms, $3s^2 3p^2$. We approximated the exchange correlation functional with the generalized-gradient approximation PBEsol \cite{PBEsol}. We used a kinetic energy cutoff for the plane wave basis of 400\,eV and a Gaussian smearing of 0.02\,eV. We used $\Gamma$-centered \textbf{k}-point grids with a \textbf{k}-spacing of $0.1$\,$\text{\AA}^{-1}$. For plotting the DOS, a denser \textbf{k}-point grid with a \textbf{k}-spacing of 0.06\,$\text{\AA}^{-1}$ and the tetrahedron method was used. All the structures were optimized until the forces are below 0.001\,eV/\AA. 
An on-site Hubbard interaction $U$ is applied to the Ce $4f$ orbitals based on the simplified rotationally invariant DFT+$U$ method by Dudarev and co-workers\cite{DFT+U}.
Unless otherwise specified, $U=9$\,eV is used, as it correctly reproduces the experimentally observed charge orders.
We performed harmonic phonon calculations using the finite displacement method in conjunction with nondiagonal supercells \cite{nondiagonal_supercells}. The dynamical matrices were calculated on a uniform \textbf{q}-point grid of size $6\times6\times6$ for the parent $P6/mmm$ structure.
\\

\textbf{Magnetotransport measurements:}
Magnetotransport measurements were carried out in a standard four-probe method using the Physical Property Measurement System (PPMS, Quantum Design). The diagonal arrangement of voltage contacts was used and the magnetoresistance and Hall resistance were obtained by symmetrization and anti-symmetrization of the measured data, respectively.\\

\textbf{Muon Spin Rotation Experiments:}

Zero and transverse-field (TF) $\mu$SR experiments were performed on the FLAME instrument using the Dilution fridge (30~mK - 2~K) and on GPS instrument using quantum cryostat (1.6 - 300~K) \cite{amato2017new}, at the Swiss Muon Source (S$\mu$S) at the Paul Scherrer Institut, in Villigen, Switzerland. High transverse-field experiments were carried out on the HAL-9500 instrument in a field range of 0.01 to 8 T. The sample in the form of pellet (diameter 6 mm) was placed on the silver sample holder and mounted in the cryostat.\\

In a ${\mu}$SR experiment nearly 100 ${\%}$ spin-polarized muons $\mu$$^{+}$ are implanted into the sample one at a time. The positively charged $\mu$$^{+}$ thermalize at interstitial lattice sites, where they act as magnetic microprobes. In a magnetic material, the muon spin precesses in the local field $B_{\rm \mu}$ at the muon site with the Larmor frequency ${\nu}_{\rm \mu}$ = $\gamma_{\rm \mu}$/(2${\pi})$$B_{\rm \mu}$, where $\gamma_{\rm \mu}$/(2${\pi}$) = 135.5 MHz T$^{-1}$ is the muon gyromagnetic ratio. Using the $\mu$SR technique, important length scales of superconductors can be measured, namely the magnetic penetration depth $\lambda$ and the coherence length $\xi$. If a type-II superconductor is cooled below $T_{\rm c}$ in an applied magnetic field between the lower ($H_{c1}$) and the upper ($H_{c2}$) critical fields, a vortex lattice is formed, which in general is incommensurate with the crystal lattice with vortex cores separated by much larger distances than those of the unit cell. Because the implanted muons stop at given crystallographic sites, they will randomly probe the field distribution of the vortex lattice. Such measurements need to be performed in a field applied perpendicular to the initial muon spin polarization (so called TF configuration).\\

\textbf{Analysis of TF-${\mu}$SR data}:

In order to model the asymmetric field distribution $P (B)$ in the SC state, the TF-${\mu}$SR time spectra measured below $T_{\rm c}$ are analyzed by using the following two-component functional form:

\begin{equation}
	\begin{aligned}
		A_{\rm TF} (t) = \sum_{i=0}^{2} A_{s,i}\exp\Big[-\frac{\sigma_{i}^2t^2}{2}\Big]\cos(\gamma_{\mu}B_{{\rm int},s,i}t+\varphi)  
		\label{eqS1}
	\end{aligned}
\end{equation}

Here $A_{s,i}$, $\sigma_{i}$ and $B_{{\rm int},s,i}$ are the initial asymmetry, relaxation rate, and local internal magnetic field of the $i$-th component, respectively. ${\varphi}$ is the initial phase of the muon-spin ensemble. $\gamma_{\mu}/(2{\pi})\simeq 135.5$~MHz/T is the muon gyromagnetic ratio. The first and second moments of the local magnetic field distribution are given by~\cite{Khasanov104504}

\begin{equation}
	\begin{aligned}
		\left\langle {B}\right\rangle = \sum_{i=0}^{2} \frac {A_{s,i}B_{{\rm int},s,i}}{A_{s,1}+A_{s,2}}
		\label{eqS2}
	\end{aligned}
\end{equation}
and 
\begin{equation}
	\begin{aligned}
		\left\langle {\Delta B}\right\rangle ^2 = \frac {\sigma ^2}{\gamma_{\mu}^2} =  \sum_{i=0}^2  \frac {A_{s,i}}{A_{s,1}+A_{s,2}} \Big[\sigma_i^2/\gamma_{\mu}^2 + \left(B_{{\rm int},s,i} - \langle B \rangle\right)^2\Big].
	\end{aligned}
\end{equation}

Above $T_{\rm c}$, in the normal state, the symmetric field distribution could be nicely modeled by only one component. The obtained relaxation rate and internal magnetic field are denoted by $\sigma_{\rm NS}$ and $B_{{\rm int},s,{\rm ns}}$. $\sigma_{\rm ns}$ is found to be small and temperature independent (dominated by nuclear magnetic moments) above $T_{\rm c}$ and assumed to be constant in the whole temperature range. Below $T_{\rm c}$, in the SC state, the relaxation rate and internal magnetic field are indicated by $\sigma_{\rm SC}$ and $B_{{\rm int},s,{\rm sc}}$. $\sigma_{\rm SC}$ is extracted by using $\sigma_\mathrm{SC} = \sqrt{\sigma^{2} - \sigma^{2}_\mathrm{ns}}$. $B_{{\rm int},s,{\rm sc}}$ is evaluated from $\left\langle {B}\right\rangle$ using Eq.~\eqref{eqS2}.\\

\textbf{Analysis of ${\lambda}(T)$}:

${\lambda}$($T$) was calculated within the local (London) approximation (${\lambda}$ ${\gg}$ ${\xi}$) by the following expression \cite{Suter69, Tinkham2004}:
\begin{equation}
	\frac{\sigma_{\rm SC}(T,\Delta_{0,i})}{\sigma_{\rm SC}(0,\Delta_{0,i})}=
	1+\frac{1}{\pi}\int_{0}^{2\pi}\int_{\Delta(_{T,\varphi})}^{\infty}(\frac{\partial f}{\partial E})\frac{EdEd\varphi}{\sqrt{E^2-\Delta_i(T,\varphi)^2}},
\end{equation}
where $f=[1+\exp(E/k_{\rm B}T)]^{-1}$ is the Fermi function, ${\varphi}$ is the angle along the Fermi surface, and ${\Delta}_{i}(T,{\varphi})={\Delta}_{0,i}{\Gamma}(T/T_{\rm c})g({\varphi}$)
(${\Delta}_{0,i}$ is the maximum gap value at $T=0$). The temperature dependence of the gap is approximated by the expression ${\Gamma}(T/T_{\rm c})=\tanh{\{}1.82[1.018(T_{\rm c}/T-1)]^{0.51}{\}}$,\cite{Carrington205} while $g({\varphi}$) describes the angular dependence of the gap and it is replaced by 1 for both an $s$-wave and an $s$+$s$-wave gap, and ${\mid}\cos(2{\varphi}){\mid}$ for a $d$-wave gap~\cite{guguchia2015direct}.\\ 

\subsection*{Acknowledgments}
This work is based on experiments performed at the Swiss Muon Source (S${\mu}$S) Paul Scherrer Insitute, Villigen, Switzerland. Z.G. acknowledges support from the Swiss National Science Foundation (SNSF) through SNSF Starting Grant (No. TMSGI2${\_}$211750).
S.-W.K. acknowledges support from a Leverhulme Trust Early Career Fellowship (ECF-2024-052) and from the research fund of Hanyang University (HY-202600000001189). B.M. and S.-W.K. acknowledge support from a UKRI Future Leaders Fellowship [MR/V023926/1]. 
The computational resources were provided by the UK National Supercomputing Service ARCHER2, for which access was obtained via the UKCP consortium and funded by EPSRC [EP/X035891/1]. Z.G. acknowledges helpful discussions with PD Dr. Mark H. Fischer.\\

\section*{Author contributions}
Z.G. conceived, designed and supervised the project. Sample growth: M.S. and F.v.R.. X-ray diffraction experiments: O.G., P.K., I.B., M.S., J.O., L.M., I.P., M.v.Z., J.C. and Z.G.. Magnetotransport experiments: O.G., P.K., M.S. and Z.G. with contributions from A.S. and N.S.. Zero and low-field $\mu$SR experiments: O.G., P.K., J.G., J.A.K., T.J.H., R.K., H.L.,  and Z.G.. High field $\mu$SR experiments: O.G., P.K., J.N.G., A.D. and Z.G.. First-principles calculations: S.W.K. and B.M., in coordination with Z.G.. Data analysis, figure development and writing of the paper: O.G. and Z.G. All authors discussed the results, interpretation, and conclusion.\\

\section*{Data availability}
The data that support the findings of this study are available from the corresponding authors upon request.\\

\section*{Conflict of Interest}
The authors declare no financial/commercial conflict of interest.\\

\bibliography{References}{}

\begin{thebibliography}{54}%
\makeatletter
\providecommand \@ifxundefined [1]{%
 \@ifx{#1\undefined}
}%
\providecommand \@ifnum [1]{%
 \ifnum #1\expandafter \@firstoftwo
 \else \expandafter \@secondoftwo
 \fi
}%
\providecommand \@ifx [1]{%
 \ifx #1\expandafter \@firstoftwo
 \else \expandafter \@secondoftwo
 \fi
}%
\providecommand \natexlab [1]{#1}%
\providecommand \enquote  [1]{``#1''}%
\providecommand \bibnamefont  [1]{#1}%
\providecommand \bibfnamefont [1]{#1}%
\providecommand \citenamefont [1]{#1}%
\providecommand \href@noop [0]{\@secondoftwo}%
\providecommand \href [0]{\begingroup \@sanitize@url \@href}%
\providecommand \@href[1]{\@@startlink{#1}\@@href}%
\providecommand \@@href[1]{\endgroup#1\@@endlink}%
\providecommand \@sanitize@url [0]{\catcode `\\12\catcode `\$12\catcode `\&12\catcode `\#12\catcode `\^12\catcode `\_12\catcode `\%12\relax}%
\providecommand \@@startlink[1]{}%
\providecommand \@@endlink[0]{}%
\providecommand \url  [0]{\begingroup\@sanitize@url \@url }%
\providecommand \@url [1]{\endgroup\@href {#1}{\urlprefix }}%
\providecommand \urlprefix  [0]{URL }%
\providecommand \Eprint [0]{\href }%
\providecommand \doibase [0]{http://dx.doi.org/}%
\providecommand \selectlanguage [0]{\@gobble}%
\providecommand \bibinfo  [0]{\@secondoftwo}%
\providecommand \bibfield  [0]{\@secondoftwo}%
\providecommand \translation [1]{[#1]}%
\providecommand \BibitemOpen [0]{}%
\providecommand \bibitemStop [0]{}%
\providecommand \bibitemNoStop [0]{.\EOS\space}%
\providecommand \EOS [0]{\spacefactor3000\relax}%
\providecommand \BibitemShut  [1]{\csname bibitem#1\endcsname}%
\let\auto@bib@innerbib\@empty
\bibitem [{\citenamefont {Kiesel}\ and\ \citenamefont {Thomale}(2012)}]{kiesel2012sublattice}%
  \BibitemOpen
  \bibfield  {author} {\bibinfo {author} {\bibfnamefont {M.~L.}\ \bibnamefont {Kiesel}}\ and\ \bibinfo {author} {\bibfnamefont {R.}~\bibnamefont {Thomale}},\ }\href@noop {} {\bibfield  {journal} {\bibinfo  {journal} {Phys. Rev. B}\ }\textbf {\bibinfo {volume} {86}},\ \bibinfo {pages} {121105} (\bibinfo {year} {2012})}\BibitemShut {NoStop}%
\bibitem [{\citenamefont {Kang}\ \emph {et~al.}(2022)\citenamefont {Kang}, \citenamefont {Fang}, \citenamefont {Kim}, \citenamefont {Ortiz}, \citenamefont {Ryu}, \citenamefont {Kim}, \citenamefont {Yoo}, \citenamefont {Sangiovanni}, \citenamefont {Di~Sante}, \citenamefont {Park}, \citenamefont {Jozwiak}, \citenamefont {Bostwick}, \citenamefont {Rotenberg}, \citenamefont {Kaxiras}, \citenamefont {Wilson}, \citenamefont {Park},\ and\ \citenamefont {Comin}}]{Kang_2022_twofold}%
  \BibitemOpen
  \bibfield  {author} {\bibinfo {author} {\bibfnamefont {M.}~\bibnamefont {Kang}}, \bibinfo {author} {\bibfnamefont {S.}~\bibnamefont {Fang}}, \bibinfo {author} {\bibfnamefont {J.-K.}\ \bibnamefont {Kim}}, \bibinfo {author} {\bibfnamefont {B.~R.}\ \bibnamefont {Ortiz}}, \bibinfo {author} {\bibfnamefont {S.~H.}\ \bibnamefont {Ryu}}, \bibinfo {author} {\bibfnamefont {J.}~\bibnamefont {Kim}}, \bibinfo {author} {\bibfnamefont {J.}~\bibnamefont {Yoo}}, \bibinfo {author} {\bibfnamefont {G.}~\bibnamefont {Sangiovanni}}, \bibinfo {author} {\bibfnamefont {D.}~\bibnamefont {Di~Sante}}, \bibinfo {author} {\bibfnamefont {B.-G.}\ \bibnamefont {Park}}, \bibinfo {author} {\bibfnamefont {C.}~\bibnamefont {Jozwiak}}, \bibinfo {author} {\bibfnamefont {A.}~\bibnamefont {Bostwick}}, \bibinfo {author} {\bibfnamefont {E.}~\bibnamefont {Rotenberg}}, \bibinfo {author} {\bibfnamefont {E.}~\bibnamefont {Kaxiras}}, \bibinfo {author} {\bibfnamefont {S.~D.}\ \bibnamefont {Wilson}}, \bibinfo {author} {\bibfnamefont {J.-H.}\ \bibnamefont
  {Park}}, \ and\ \bibinfo {author} {\bibfnamefont {R.}~\bibnamefont {Comin}},\ }\href {\doibase 10.1038/s41567-021-01451-5} {\bibfield  {journal} {\bibinfo  {journal} {Nat. Phys.}\ }\textbf {\bibinfo {volume} {18}},\ \bibinfo {pages} {301} (\bibinfo {year} {2022})}\BibitemShut {NoStop}%
\bibitem [{\citenamefont {Kim}\ \emph {et~al.}(2023)\citenamefont {Kim}, \citenamefont {Oh}, \citenamefont {Moon},\ and\ \citenamefont {Kim}}]{Kim2023_monolayer}%
  \BibitemOpen
  \bibfield  {author} {\bibinfo {author} {\bibfnamefont {S.-W.}\ \bibnamefont {Kim}}, \bibinfo {author} {\bibfnamefont {H.}~\bibnamefont {Oh}}, \bibinfo {author} {\bibfnamefont {E.-G.}\ \bibnamefont {Moon}}, \ and\ \bibinfo {author} {\bibfnamefont {Y.}~\bibnamefont {Kim}},\ }\href {\doibase 10.1038/s41467-023-36341-2} {\bibfield  {journal} {\bibinfo  {journal} {Nat. Commun.}\ }\textbf {\bibinfo {volume} {14}},\ \bibinfo {pages} {591} (\bibinfo {year} {2023})}\BibitemShut {NoStop}%
\bibitem [{\citenamefont {Lin}\ \emph {et~al.}(2024)\citenamefont {Lin}, \citenamefont {Consiglio}, \citenamefont {Forslund}, \citenamefont {K{\"u}spert}, \citenamefont {Denner}, \citenamefont {Lei}, \citenamefont {Louat}, \citenamefont {Watson}, \citenamefont {Kim}, \citenamefont {Cacho} \emph {et~al.}}]{lin2024uniaxial}%
  \BibitemOpen
  \bibfield  {author} {\bibinfo {author} {\bibfnamefont {C.}~\bibnamefont {Lin}}, \bibinfo {author} {\bibfnamefont {A.}~\bibnamefont {Consiglio}}, \bibinfo {author} {\bibfnamefont {O.~K.}\ \bibnamefont {Forslund}}, \bibinfo {author} {\bibfnamefont {J.}~\bibnamefont {K{\"u}spert}}, \bibinfo {author} {\bibfnamefont {M.~M.}\ \bibnamefont {Denner}}, \bibinfo {author} {\bibfnamefont {H.}~\bibnamefont {Lei}}, \bibinfo {author} {\bibfnamefont {A.}~\bibnamefont {Louat}}, \bibinfo {author} {\bibfnamefont {M.~D.}\ \bibnamefont {Watson}}, \bibinfo {author} {\bibfnamefont {T.~K.}\ \bibnamefont {Kim}}, \bibinfo {author} {\bibfnamefont {C.}~\bibnamefont {Cacho}},  \emph {et~al.},\ }\href@noop {} {\bibfield  {journal} {\bibinfo  {journal} {Nat. Commun.}\ }\textbf {\bibinfo {volume} {15}},\ \bibinfo {pages} {10466} (\bibinfo {year} {2024})}\BibitemShut {NoStop}%
\bibitem [{\citenamefont {Wang}\ \emph {et~al.}(2023)\citenamefont {Wang}, \citenamefont {Wu}, \citenamefont {McCandless}, \citenamefont {Chan},\ and\ \citenamefont {Ali}}]{wang2023quantum}%
  \BibitemOpen
  \bibfield  {author} {\bibinfo {author} {\bibfnamefont {Y.}~\bibnamefont {Wang}}, \bibinfo {author} {\bibfnamefont {H.}~\bibnamefont {Wu}}, \bibinfo {author} {\bibfnamefont {G.~T.}\ \bibnamefont {McCandless}}, \bibinfo {author} {\bibfnamefont {J.~Y.}\ \bibnamefont {Chan}}, \ and\ \bibinfo {author} {\bibfnamefont {M.~N.}\ \bibnamefont {Ali}},\ }\href@noop {} {\bibfield  {journal} {\bibinfo  {journal} {Nat. Rev. Phys.}\ }\textbf {\bibinfo {volume} {5}},\ \bibinfo {pages} {635} (\bibinfo {year} {2023})}\BibitemShut {NoStop}%
\bibitem [{\citenamefont {Mielke~III}\ \emph {et~al.}(2021)\citenamefont {Mielke~III}, \citenamefont {Qin}, \citenamefont {Yin}, \citenamefont {Nakamura}, \citenamefont {Das}, \citenamefont {Guo}, \citenamefont {Khasanov}, \citenamefont {Chang}, \citenamefont {Wang}, \citenamefont {Jia} \emph {et~al.}}]{mielke2021nodeless}%
  \BibitemOpen
  \bibfield  {author} {\bibinfo {author} {\bibfnamefont {C.}~\bibnamefont {Mielke~III}}, \bibinfo {author} {\bibfnamefont {Y.}~\bibnamefont {Qin}}, \bibinfo {author} {\bibfnamefont {J.-X.}\ \bibnamefont {Yin}}, \bibinfo {author} {\bibfnamefont {H.}~\bibnamefont {Nakamura}}, \bibinfo {author} {\bibfnamefont {D.}~\bibnamefont {Das}}, \bibinfo {author} {\bibfnamefont {K.}~\bibnamefont {Guo}}, \bibinfo {author} {\bibfnamefont {R.}~\bibnamefont {Khasanov}}, \bibinfo {author} {\bibfnamefont {J.}~\bibnamefont {Chang}}, \bibinfo {author} {\bibfnamefont {Z.}~\bibnamefont {Wang}}, \bibinfo {author} {\bibfnamefont {S.}~\bibnamefont {Jia}},  \emph {et~al.},\ }\href@noop {} {\bibfield  {journal} {\bibinfo  {journal} {Phys. Rev. Mater.}\ }\textbf {\bibinfo {volume} {5}},\ \bibinfo {pages} {034803} (\bibinfo {year} {2021})}\BibitemShut {NoStop}%
\bibitem [{\citenamefont {Guguchia}\ \emph {et~al.}(2023{\natexlab{a}})\citenamefont {Guguchia}, \citenamefont {Mielke~III}, \citenamefont {Das}, \citenamefont {Gupta}, \citenamefont {Yin}, \citenamefont {Liu}, \citenamefont {Yin}, \citenamefont {Christensen}, \citenamefont {Tu}, \citenamefont {Gong} \emph {et~al.}}]{guguchia2023tunable}%
  \BibitemOpen
  \bibfield  {author} {\bibinfo {author} {\bibfnamefont {Z.}~\bibnamefont {Guguchia}}, \bibinfo {author} {\bibfnamefont {C.}~\bibnamefont {Mielke~III}}, \bibinfo {author} {\bibfnamefont {D.}~\bibnamefont {Das}}, \bibinfo {author} {\bibfnamefont {R.}~\bibnamefont {Gupta}}, \bibinfo {author} {\bibfnamefont {J.-X.}\ \bibnamefont {Yin}}, \bibinfo {author} {\bibfnamefont {H.}~\bibnamefont {Liu}}, \bibinfo {author} {\bibfnamefont {Q.}~\bibnamefont {Yin}}, \bibinfo {author} {\bibfnamefont {M.~H.}\ \bibnamefont {Christensen}}, \bibinfo {author} {\bibfnamefont {Z.}~\bibnamefont {Tu}}, \bibinfo {author} {\bibfnamefont {C.}~\bibnamefont {Gong}},  \emph {et~al.},\ }\href@noop {} {\bibfield  {journal} {\bibinfo  {journal} {Nat. Commun.}\ }\textbf {\bibinfo {volume} {14}},\ \bibinfo {pages} {153} (\bibinfo {year} {2023}{\natexlab{a}})}\BibitemShut {NoStop}%
\bibitem [{\citenamefont {Jovanovic}\ and\ \citenamefont {Schoop}(2022)}]{jovanovic2022simple}%
  \BibitemOpen
  \bibfield  {author} {\bibinfo {author} {\bibfnamefont {M.}~\bibnamefont {Jovanovic}}\ and\ \bibinfo {author} {\bibfnamefont {L.~M.}\ \bibnamefont {Schoop}},\ }\href@noop {} {\bibfield  {journal} {\bibinfo  {journal} {J. Am. Chem. Soc.}\ }\textbf {\bibinfo {volume} {144}},\ \bibinfo {pages} {10978} (\bibinfo {year} {2022})}\BibitemShut {NoStop}%
\bibitem [{\citenamefont {Kang}\ \emph {et~al.}(2020)\citenamefont {Kang}, \citenamefont {Ye}, \citenamefont {Fang}, \citenamefont {You}, \citenamefont {Levitan}, \citenamefont {Han}, \citenamefont {Facio}, \citenamefont {Jozwiak}, \citenamefont {Bostwick}, \citenamefont {Rotenberg} \emph {et~al.}}]{kang2020dirac}%
  \BibitemOpen
  \bibfield  {author} {\bibinfo {author} {\bibfnamefont {M.}~\bibnamefont {Kang}}, \bibinfo {author} {\bibfnamefont {L.}~\bibnamefont {Ye}}, \bibinfo {author} {\bibfnamefont {S.}~\bibnamefont {Fang}}, \bibinfo {author} {\bibfnamefont {J.-S.}\ \bibnamefont {You}}, \bibinfo {author} {\bibfnamefont {A.}~\bibnamefont {Levitan}}, \bibinfo {author} {\bibfnamefont {M.}~\bibnamefont {Han}}, \bibinfo {author} {\bibfnamefont {J.~I.}\ \bibnamefont {Facio}}, \bibinfo {author} {\bibfnamefont {C.}~\bibnamefont {Jozwiak}}, \bibinfo {author} {\bibfnamefont {A.}~\bibnamefont {Bostwick}}, \bibinfo {author} {\bibfnamefont {E.}~\bibnamefont {Rotenberg}},  \emph {et~al.},\ }\href@noop {} {\bibfield  {journal} {\bibinfo  {journal} {Nat. Mater.}\ }\textbf {\bibinfo {volume} {19}},\ \bibinfo {pages} {163} (\bibinfo {year} {2020})}\BibitemShut {NoStop}%
\bibitem [{\citenamefont {Gerguri}\ \emph {et~al.}(2025)\citenamefont {Gerguri}, \citenamefont {Das}, \citenamefont {Sazgari}, \citenamefont {Liu}, \citenamefont {Mielke~III}, \citenamefont {Kr{\`a}l}, \citenamefont {Islam}, \citenamefont {Graham}, \citenamefont {Grinenko}, \citenamefont {Sarkar} \emph {et~al.}}]{gerguri2025distinct}%
  \BibitemOpen
  \bibfield  {author} {\bibinfo {author} {\bibfnamefont {O.}~\bibnamefont {Gerguri}}, \bibinfo {author} {\bibfnamefont {D.}~\bibnamefont {Das}}, \bibinfo {author} {\bibfnamefont {V.}~\bibnamefont {Sazgari}}, \bibinfo {author} {\bibfnamefont {H.}~\bibnamefont {Liu}}, \bibinfo {author} {\bibfnamefont {C.}~\bibnamefont {Mielke~III}}, \bibinfo {author} {\bibfnamefont {P.}~\bibnamefont {Kr{\`a}l}}, \bibinfo {author} {\bibfnamefont {S.}~\bibnamefont {Islam}}, \bibinfo {author} {\bibfnamefont {J.}~\bibnamefont {Graham}}, \bibinfo {author} {\bibfnamefont {V.}~\bibnamefont {Grinenko}}, \bibinfo {author} {\bibfnamefont {R.}~\bibnamefont {Sarkar}},  \emph {et~al.},\ }\href@noop {} {\bibfield  {journal} {\bibinfo  {journal} {arXiv preprint arXiv:2507.09779}\ } (\bibinfo {year} {2025})}\BibitemShut {NoStop}%
\bibitem [{\citenamefont {Ye}\ \emph {et~al.}(2018)\citenamefont {Ye}, \citenamefont {Kang}, \citenamefont {Liu}, \citenamefont {Von~Cube}, \citenamefont {Wicker}, \citenamefont {Suzuki}, \citenamefont {Jozwiak}, \citenamefont {Bostwick}, \citenamefont {Rotenberg}, \citenamefont {Bell} \emph {et~al.}}]{ye2018massive}%
  \BibitemOpen
  \bibfield  {author} {\bibinfo {author} {\bibfnamefont {L.}~\bibnamefont {Ye}}, \bibinfo {author} {\bibfnamefont {M.}~\bibnamefont {Kang}}, \bibinfo {author} {\bibfnamefont {J.}~\bibnamefont {Liu}}, \bibinfo {author} {\bibfnamefont {F.}~\bibnamefont {Von~Cube}}, \bibinfo {author} {\bibfnamefont {C.~R.}\ \bibnamefont {Wicker}}, \bibinfo {author} {\bibfnamefont {T.}~\bibnamefont {Suzuki}}, \bibinfo {author} {\bibfnamefont {C.}~\bibnamefont {Jozwiak}}, \bibinfo {author} {\bibfnamefont {A.}~\bibnamefont {Bostwick}}, \bibinfo {author} {\bibfnamefont {E.}~\bibnamefont {Rotenberg}}, \bibinfo {author} {\bibfnamefont {D.~C.}\ \bibnamefont {Bell}},  \emph {et~al.},\ }\href@noop {} {\bibfield  {journal} {\bibinfo  {journal} {Nature}\ }\textbf {\bibinfo {volume} {555}},\ \bibinfo {pages} {638} (\bibinfo {year} {2018})}\BibitemShut {NoStop}%
\bibitem [{\citenamefont {Chen}\ \emph {et~al.}(2021)\citenamefont {Chen}, \citenamefont {Yang}, \citenamefont {Hu}, \citenamefont {Zhao}, \citenamefont {Yuan}, \citenamefont {Xing}, \citenamefont {Qian}, \citenamefont {Huang}, \citenamefont {Li}, \citenamefont {Ye} \emph {et~al.}}]{chen2021roton}%
  \BibitemOpen
  \bibfield  {author} {\bibinfo {author} {\bibfnamefont {H.}~\bibnamefont {Chen}}, \bibinfo {author} {\bibfnamefont {H.}~\bibnamefont {Yang}}, \bibinfo {author} {\bibfnamefont {B.}~\bibnamefont {Hu}}, \bibinfo {author} {\bibfnamefont {Z.}~\bibnamefont {Zhao}}, \bibinfo {author} {\bibfnamefont {J.}~\bibnamefont {Yuan}}, \bibinfo {author} {\bibfnamefont {Y.}~\bibnamefont {Xing}}, \bibinfo {author} {\bibfnamefont {G.}~\bibnamefont {Qian}}, \bibinfo {author} {\bibfnamefont {Z.}~\bibnamefont {Huang}}, \bibinfo {author} {\bibfnamefont {G.}~\bibnamefont {Li}}, \bibinfo {author} {\bibfnamefont {Y.}~\bibnamefont {Ye}},  \emph {et~al.},\ }\href@noop {} {\bibfield  {journal} {\bibinfo  {journal} {Nature}\ }\textbf {\bibinfo {volume} {599}},\ \bibinfo {pages} {222} (\bibinfo {year} {2021})}\BibitemShut {NoStop}%
\bibitem [{\citenamefont {Guguchia}\ \emph {et~al.}(2023{\natexlab{b}})\citenamefont {Guguchia}, \citenamefont {Khasanov},\ and\ \citenamefont {Luetkens}}]{guguchia2023unconventional}%
  \BibitemOpen
  \bibfield  {author} {\bibinfo {author} {\bibfnamefont {Z.}~\bibnamefont {Guguchia}}, \bibinfo {author} {\bibfnamefont {R.}~\bibnamefont {Khasanov}}, \ and\ \bibinfo {author} {\bibfnamefont {H.}~\bibnamefont {Luetkens}},\ }\href@noop {} {\bibfield  {journal} {\bibinfo  {journal} {npj Quantum Mater.}\ }\textbf {\bibinfo {volume} {8}},\ \bibinfo {pages} {41} (\bibinfo {year} {2023}{\natexlab{b}})}\BibitemShut {NoStop}%
\bibitem [{\citenamefont {Mielke~III}\ \emph {et~al.}(2022)\citenamefont {Mielke~III}, \citenamefont {Das}, \citenamefont {Yin}, \citenamefont {Liu}, \citenamefont {Gupta}, \citenamefont {Jiang}, \citenamefont {Medarde}, \citenamefont {Wu}, \citenamefont {Lei}, \citenamefont {Chang} \emph {et~al.}}]{mielke2022time}%
  \BibitemOpen
  \bibfield  {author} {\bibinfo {author} {\bibfnamefont {C.}~\bibnamefont {Mielke~III}}, \bibinfo {author} {\bibfnamefont {D.}~\bibnamefont {Das}}, \bibinfo {author} {\bibfnamefont {J.-X.}\ \bibnamefont {Yin}}, \bibinfo {author} {\bibfnamefont {H.}~\bibnamefont {Liu}}, \bibinfo {author} {\bibfnamefont {R.}~\bibnamefont {Gupta}}, \bibinfo {author} {\bibfnamefont {Y.-X.}\ \bibnamefont {Jiang}}, \bibinfo {author} {\bibfnamefont {M.}~\bibnamefont {Medarde}}, \bibinfo {author} {\bibfnamefont {X.}~\bibnamefont {Wu}}, \bibinfo {author} {\bibfnamefont {H.~C.}\ \bibnamefont {Lei}}, \bibinfo {author} {\bibfnamefont {J.}~\bibnamefont {Chang}},  \emph {et~al.},\ }\href@noop {} {\bibfield  {journal} {\bibinfo  {journal} {Nature}\ }\textbf {\bibinfo {volume} {602}},\ \bibinfo {pages} {245} (\bibinfo {year} {2022})}\BibitemShut {NoStop}%
\bibitem [{\citenamefont {Jiang}\ \emph {et~al.}(2021)\citenamefont {Jiang}, \citenamefont {Yin}, \citenamefont {Denner}, \citenamefont {Shumiya}, \citenamefont {Ortiz}, \citenamefont {Xu}, \citenamefont {Guguchia}, \citenamefont {He}, \citenamefont {Hossain}, \citenamefont {Liu} \emph {et~al.}}]{jiang2021unconventional}%
  \BibitemOpen
  \bibfield  {author} {\bibinfo {author} {\bibfnamefont {Y.-X.}\ \bibnamefont {Jiang}}, \bibinfo {author} {\bibfnamefont {J.-X.}\ \bibnamefont {Yin}}, \bibinfo {author} {\bibfnamefont {M.~M.}\ \bibnamefont {Denner}}, \bibinfo {author} {\bibfnamefont {N.}~\bibnamefont {Shumiya}}, \bibinfo {author} {\bibfnamefont {B.~R.}\ \bibnamefont {Ortiz}}, \bibinfo {author} {\bibfnamefont {G.}~\bibnamefont {Xu}}, \bibinfo {author} {\bibfnamefont {Z.}~\bibnamefont {Guguchia}}, \bibinfo {author} {\bibfnamefont {J.}~\bibnamefont {He}}, \bibinfo {author} {\bibfnamefont {M.~S.}\ \bibnamefont {Hossain}}, \bibinfo {author} {\bibfnamefont {X.}~\bibnamefont {Liu}},  \emph {et~al.},\ }\href@noop {} {\bibfield  {journal} {\bibinfo  {journal} {Nat. Mater.}\ }\textbf {\bibinfo {volume} {20}},\ \bibinfo {pages} {1353} (\bibinfo {year} {2021})}\BibitemShut {NoStop}%
\bibitem [{\citenamefont {Mielke~III}\ \emph {et~al.}(2025)\citenamefont {Mielke~III}, \citenamefont {Sazgari}, \citenamefont {Plokhikh}, \citenamefont {Yi}, \citenamefont {Shin}, \citenamefont {Nakamura}, \citenamefont {Graham}, \citenamefont {K{\"u}spert}, \citenamefont {Bia{\l}o}, \citenamefont {Garbarino} \emph {et~al.}}]{mielke2025coexisting}%
  \BibitemOpen
  \bibfield  {author} {\bibinfo {author} {\bibfnamefont {C.}~\bibnamefont {Mielke~III}}, \bibinfo {author} {\bibfnamefont {V.}~\bibnamefont {Sazgari}}, \bibinfo {author} {\bibfnamefont {I.}~\bibnamefont {Plokhikh}}, \bibinfo {author} {\bibfnamefont {M.}~\bibnamefont {Yi}}, \bibinfo {author} {\bibfnamefont {S.}~\bibnamefont {Shin}}, \bibinfo {author} {\bibfnamefont {H.}~\bibnamefont {Nakamura}}, \bibinfo {author} {\bibfnamefont {J.}~\bibnamefont {Graham}}, \bibinfo {author} {\bibfnamefont {J.}~\bibnamefont {K{\"u}spert}}, \bibinfo {author} {\bibfnamefont {I.}~\bibnamefont {Bia{\l}o}}, \bibinfo {author} {\bibfnamefont {G.}~\bibnamefont {Garbarino}},  \emph {et~al.},\ }\href@noop {} {\bibfield  {journal} {\bibinfo  {journal} {Adv. Mater.}\ }\textbf {\bibinfo {volume} {37}},\ \bibinfo {pages} {2503065} (\bibinfo {year} {2025})}\BibitemShut {NoStop}%
\bibitem [{\citenamefont {Huang}\ \emph {et~al.}(2026)\citenamefont {Huang}, \citenamefont {Ren}, \citenamefont {Tan}, \citenamefont {Hyun}, \citenamefont {Zhang}, \citenamefont {Hulse}, \citenamefont {Liu}, \citenamefont {DeStefano}, \citenamefont {Xie}, \citenamefont {Yue} \emph {et~al.}}]{huang2026magnetic}%
  \BibitemOpen
  \bibfield  {author} {\bibinfo {author} {\bibfnamefont {J.}~\bibnamefont {Huang}}, \bibinfo {author} {\bibfnamefont {Z.}~\bibnamefont {Ren}}, \bibinfo {author} {\bibfnamefont {H.}~\bibnamefont {Tan}}, \bibinfo {author} {\bibfnamefont {J.}~\bibnamefont {Hyun}}, \bibinfo {author} {\bibfnamefont {Y.}~\bibnamefont {Zhang}}, \bibinfo {author} {\bibfnamefont {T.~A.}\ \bibnamefont {Hulse}}, \bibinfo {author} {\bibfnamefont {Z.}~\bibnamefont {Liu}}, \bibinfo {author} {\bibfnamefont {J.~M.}\ \bibnamefont {DeStefano}}, \bibinfo {author} {\bibfnamefont {Y.}~\bibnamefont {Xie}}, \bibinfo {author} {\bibfnamefont {Z.}~\bibnamefont {Yue}},  \emph {et~al.},\ }\href@noop {} {\bibfield  {journal} {\bibinfo  {journal} {Nat. Phys.}\ ,\ \bibinfo {pages} {1}} (\bibinfo {year} {2026})}\BibitemShut {NoStop}%
\bibitem [{\citenamefont {Plokhikh}\ \emph {et~al.}(2024)\citenamefont {Plokhikh}, \citenamefont {Mielke~Iii}, \citenamefont {Nakamura}, \citenamefont {Petricek}, \citenamefont {Qin}, \citenamefont {Sazgari}, \citenamefont {K{\"u}spert}, \citenamefont {Bia{\l}o}, \citenamefont {Shin}, \citenamefont {Ivashko} \emph {et~al.}}]{plokhikh2024discovery}%
  \BibitemOpen
  \bibfield  {author} {\bibinfo {author} {\bibfnamefont {I.}~\bibnamefont {Plokhikh}}, \bibinfo {author} {\bibfnamefont {C.}~\bibnamefont {Mielke~Iii}}, \bibinfo {author} {\bibfnamefont {H.}~\bibnamefont {Nakamura}}, \bibinfo {author} {\bibfnamefont {V.}~\bibnamefont {Petricek}}, \bibinfo {author} {\bibfnamefont {Y.}~\bibnamefont {Qin}}, \bibinfo {author} {\bibfnamefont {V.}~\bibnamefont {Sazgari}}, \bibinfo {author} {\bibfnamefont {J.}~\bibnamefont {K{\"u}spert}}, \bibinfo {author} {\bibfnamefont {I.}~\bibnamefont {Bia{\l}o}}, \bibinfo {author} {\bibfnamefont {S.}~\bibnamefont {Shin}}, \bibinfo {author} {\bibfnamefont {O.}~\bibnamefont {Ivashko}},  \emph {et~al.},\ }\href@noop {} {\bibfield  {journal} {\bibinfo  {journal} {Commun. Phys.}\ }\textbf {\bibinfo {volume} {7}},\ \bibinfo {pages} {182} (\bibinfo {year} {2024})}\BibitemShut {NoStop}%
\bibitem [{\citenamefont {Ma}\ \emph {et~al.}(2025)\citenamefont {Ma}, \citenamefont {Plokhikh}, \citenamefont {Graham}, \citenamefont {Mielke~III}, \citenamefont {Sazgari}, \citenamefont {Nakamura}, \citenamefont {Islam}, \citenamefont {Shin}, \citenamefont {Kr{\'a}l}, \citenamefont {Gerguri} \emph {et~al.}}]{ma2025correlation}%
  \BibitemOpen
  \bibfield  {author} {\bibinfo {author} {\bibfnamefont {K.}~\bibnamefont {Ma}}, \bibinfo {author} {\bibfnamefont {I.}~\bibnamefont {Plokhikh}}, \bibinfo {author} {\bibfnamefont {J.~N.}\ \bibnamefont {Graham}}, \bibinfo {author} {\bibfnamefont {C.}~\bibnamefont {Mielke~III}}, \bibinfo {author} {\bibfnamefont {V.}~\bibnamefont {Sazgari}}, \bibinfo {author} {\bibfnamefont {H.}~\bibnamefont {Nakamura}}, \bibinfo {author} {\bibfnamefont {S.~S.}\ \bibnamefont {Islam}}, \bibinfo {author} {\bibfnamefont {S.}~\bibnamefont {Shin}}, \bibinfo {author} {\bibfnamefont {P.}~\bibnamefont {Kr{\'a}l}}, \bibinfo {author} {\bibfnamefont {O.}~\bibnamefont {Gerguri}},  \emph {et~al.},\ }\href@noop {} {\bibfield  {journal} {\bibinfo  {journal} {Nat. Commun.}\ }\textbf {\bibinfo {volume} {16}},\ \bibinfo {pages} {6149} (\bibinfo {year} {2025})}\BibitemShut {NoStop}%
\bibitem [{\citenamefont {Li}\ \emph {et~al.}(2025)\citenamefont {Li}, \citenamefont {Huyan}, \citenamefont {Thompson}, \citenamefont {Slade}, \citenamefont {Zhang}, \citenamefont {Ryu}, \citenamefont {Bi}, \citenamefont {Bud'ko},\ and\ \citenamefont {Canfield}}]{li2025superconducting}%
  \BibitemOpen
  \bibfield  {author} {\bibinfo {author} {\bibfnamefont {Z.}~\bibnamefont {Li}}, \bibinfo {author} {\bibfnamefont {S.}~\bibnamefont {Huyan}}, \bibinfo {author} {\bibfnamefont {E.~C.}\ \bibnamefont {Thompson}}, \bibinfo {author} {\bibfnamefont {T.~J.}\ \bibnamefont {Slade}}, \bibinfo {author} {\bibfnamefont {D.}~\bibnamefont {Zhang}}, \bibinfo {author} {\bibfnamefont {Y.~J.}\ \bibnamefont {Ryu}}, \bibinfo {author} {\bibfnamefont {W.}~\bibnamefont {Bi}}, \bibinfo {author} {\bibfnamefont {S.~L.}\ \bibnamefont {Bud'ko}}, \ and\ \bibinfo {author} {\bibfnamefont {P.~C.}\ \bibnamefont {Canfield}},\ }\href@noop {} {\bibfield  {journal} {\bibinfo  {journal} {Physical Review B}\ }\textbf {\bibinfo {volume} {111}},\ \bibinfo {pages} {144505} (\bibinfo {year} {2025})}\BibitemShut {NoStop}%
\bibitem [{\citenamefont {Deng}\ \emph {et~al.}(2025)\citenamefont {Deng}, \citenamefont {Jiang}, \citenamefont {Cerqueira}, \citenamefont {Hu}, \citenamefont {Lamponen}, \citenamefont {C{\u{a}}lug{\u{a}}ru}, \citenamefont {Pi}, \citenamefont {Wang}, \citenamefont {Vergniory}, \citenamefont {Morosan} \emph {et~al.}}]{deng2025theory}%
  \BibitemOpen
  \bibfield  {author} {\bibinfo {author} {\bibfnamefont {J.}~\bibnamefont {Deng}}, \bibinfo {author} {\bibfnamefont {Y.}~\bibnamefont {Jiang}}, \bibinfo {author} {\bibfnamefont {T.~F.}\ \bibnamefont {Cerqueira}}, \bibinfo {author} {\bibfnamefont {H.}~\bibnamefont {Hu}}, \bibinfo {author} {\bibfnamefont {E.~O.}\ \bibnamefont {Lamponen}}, \bibinfo {author} {\bibfnamefont {D.}~\bibnamefont {C{\u{a}}lug{\u{a}}ru}}, \bibinfo {author} {\bibfnamefont {H.}~\bibnamefont {Pi}}, \bibinfo {author} {\bibfnamefont {Z.}~\bibnamefont {Wang}}, \bibinfo {author} {\bibfnamefont {M.~G.}\ \bibnamefont {Vergniory}}, \bibinfo {author} {\bibfnamefont {E.}~\bibnamefont {Morosan}},  \emph {et~al.},\ }\href@noop {} {\bibfield  {journal} {\bibinfo  {journal} {arXiv:2503.20867}\ } (\bibinfo {year} {2025})}\BibitemShut {NoStop}%
\bibitem [{\citenamefont {Kr{\'a}l}\ \emph {et~al.}(2026{\natexlab{a}})\citenamefont {Kr{\'a}l}, \citenamefont {Sazgari}, \citenamefont {Ge}, \citenamefont {Gerguri}, \citenamefont {Spitaler}, \citenamefont {Graham}, \citenamefont {Nakamura}, \citenamefont {Bartkowiak}, \citenamefont {Nakatsuji}, \citenamefont {Luetkens} \emph {et~al.}}]{kral2026uniaxial}%
  \BibitemOpen
  \bibfield  {author} {\bibinfo {author} {\bibfnamefont {P.}~\bibnamefont {Kr{\'a}l}}, \bibinfo {author} {\bibfnamefont {V.}~\bibnamefont {Sazgari}}, \bibinfo {author} {\bibfnamefont {Y.}~\bibnamefont {Ge}}, \bibinfo {author} {\bibfnamefont {O.}~\bibnamefont {Gerguri}}, \bibinfo {author} {\bibfnamefont {M.}~\bibnamefont {Spitaler}}, \bibinfo {author} {\bibfnamefont {J.}~\bibnamefont {Graham}}, \bibinfo {author} {\bibfnamefont {H.}~\bibnamefont {Nakamura}}, \bibinfo {author} {\bibfnamefont {M.}~\bibnamefont {Bartkowiak}}, \bibinfo {author} {\bibfnamefont {S.}~\bibnamefont {Nakatsuji}}, \bibinfo {author} {\bibfnamefont {H.}~\bibnamefont {Luetkens}},  \emph {et~al.},\ }\href@noop {} {\bibfield  {journal} {\bibinfo  {journal} {arXiv preprint arXiv:2602.15998}\ } (\bibinfo {year} {2026}{\natexlab{a}})}\BibitemShut {NoStop}%
\bibitem [{\citenamefont {Kr{\'a}l}\ \emph {et~al.}(2026{\natexlab{b}})\citenamefont {Kr{\'a}l}, \citenamefont {Graham}, \citenamefont {Sazgari}, \citenamefont {Plokhikh}, \citenamefont {Lukovkina}, \citenamefont {Gerguri}, \citenamefont {Bia{\l}o}, \citenamefont {Doll}, \citenamefont {Martinelli}, \citenamefont {Oppliger}, \citenamefont {Islam}, \citenamefont {Spitaler}, \citenamefont {Wang}, \citenamefont {Salamin}, \citenamefont {Luetkens}, \citenamefont {Khasanov}, \citenamefont {Zimmermann}, \citenamefont {Yin}, \citenamefont {Wang}, \citenamefont {Chang}, \citenamefont {Monserrat}, \citenamefont {Gawryluk}, \citenamefont {von Rohr}, \citenamefont {Kim},\ and\ \citenamefont {Guguchia}}]{kral2026chargeorder}%
  \BibitemOpen
  \bibfield  {author} {\bibinfo {author} {\bibfnamefont {P.}~\bibnamefont {Kr{\'a}l}}, \bibinfo {author} {\bibfnamefont {J.~N.}\ \bibnamefont {Graham}}, \bibinfo {author} {\bibfnamefont {V.}~\bibnamefont {Sazgari}}, \bibinfo {author} {\bibfnamefont {I.}~\bibnamefont {Plokhikh}}, \bibinfo {author} {\bibfnamefont {A.}~\bibnamefont {Lukovkina}}, \bibinfo {author} {\bibfnamefont {O.}~\bibnamefont {Gerguri}}, \bibinfo {author} {\bibfnamefont {I.}~\bibnamefont {Bia{\l}o}}, \bibinfo {author} {\bibfnamefont {A.}~\bibnamefont {Doll}}, \bibinfo {author} {\bibfnamefont {L.}~\bibnamefont {Martinelli}}, \bibinfo {author} {\bibfnamefont {J.}~\bibnamefont {Oppliger}}, \bibinfo {author} {\bibfnamefont {S.~S.}\ \bibnamefont {Islam}}, \bibinfo {author} {\bibfnamefont {M.}~\bibnamefont {Spitaler}}, \bibinfo {author} {\bibfnamefont {K.}~\bibnamefont {Wang}}, \bibinfo {author} {\bibfnamefont {M.}~\bibnamefont {Salamin}}, \bibinfo {author} {\bibfnamefont {H.}~\bibnamefont {Luetkens}}, \bibinfo {author} {\bibfnamefont
  {R.}~\bibnamefont {Khasanov}}, \bibinfo {author} {\bibfnamefont {M.~V.}\ \bibnamefont {Zimmermann}}, \bibinfo {author} {\bibfnamefont {J.-X.}\ \bibnamefont {Yin}}, \bibinfo {author} {\bibfnamefont {Z.}~\bibnamefont {Wang}}, \bibinfo {author} {\bibfnamefont {J.}~\bibnamefont {Chang}}, \bibinfo {author} {\bibfnamefont {B.}~\bibnamefont {Monserrat}}, \bibinfo {author} {\bibfnamefont {D.}~\bibnamefont {Gawryluk}}, \bibinfo {author} {\bibfnamefont {F.~O.}\ \bibnamefont {von Rohr}}, \bibinfo {author} {\bibfnamefont {S.-W.}\ \bibnamefont {Kim}}, \ and\ \bibinfo {author} {\bibfnamefont {Z.}~\bibnamefont {Guguchia}},\ }\href {\doibase 10.1038/s41467-025-67881-4} {\bibfield  {journal} {\bibinfo  {journal} {Nat. Commun.}\ }\textbf {\bibinfo {volume} {17}},\ \bibinfo {pages} {1121} (\bibinfo {year} {2026}{\natexlab{b}})}\BibitemShut {NoStop}%
\bibitem [{\citenamefont {Li}\ \emph {et~al.}(2016)\citenamefont {Li}, \citenamefont {Li},\ and\ \citenamefont {Wen}}]{li2016chemical}%
  \BibitemOpen
  \bibfield  {author} {\bibinfo {author} {\bibfnamefont {B.}~\bibnamefont {Li}}, \bibinfo {author} {\bibfnamefont {S.}~\bibnamefont {Li}}, \ and\ \bibinfo {author} {\bibfnamefont {H.-H.}\ \bibnamefont {Wen}},\ }\href@noop {} {\bibfield  {journal} {\bibinfo  {journal} {Phys. Rev. B}\ }\textbf {\bibinfo {volume} {94}},\ \bibinfo {pages} {094523} (\bibinfo {year} {2016})}\BibitemShut {NoStop}%
\bibitem [{\citenamefont {Rauchschwalbe}\ \emph {et~al.}(1985)\citenamefont {Rauchschwalbe}, \citenamefont {Gottwick}, \citenamefont {Ahlheim}, \citenamefont {Mayer},\ and\ \citenamefont {Steglich}}]{rauchschwalbe1985investigation}%
  \BibitemOpen
  \bibfield  {author} {\bibinfo {author} {\bibfnamefont {U.}~\bibnamefont {Rauchschwalbe}}, \bibinfo {author} {\bibfnamefont {U.}~\bibnamefont {Gottwick}}, \bibinfo {author} {\bibfnamefont {U.}~\bibnamefont {Ahlheim}}, \bibinfo {author} {\bibfnamefont {H.}~\bibnamefont {Mayer}}, \ and\ \bibinfo {author} {\bibfnamefont {F.}~\bibnamefont {Steglich}},\ }\href@noop {} {\bibfield  {journal} {\bibinfo  {journal} {Journal of the Less Common Metals}\ }\textbf {\bibinfo {volume} {111}},\ \bibinfo {pages} {265} (\bibinfo {year} {1985})}\BibitemShut {NoStop}%
\bibitem [{\citenamefont {Rauchschwalbe}\ \emph {et~al.}(1984)\citenamefont {Rauchschwalbe}, \citenamefont {Lieke}, \citenamefont {Steglich}, \citenamefont {Godart}, \citenamefont {Gupta},\ and\ \citenamefont {Parks}}]{rauchschwalbe1984superconductivity}%
  \BibitemOpen
  \bibfield  {author} {\bibinfo {author} {\bibfnamefont {U.}~\bibnamefont {Rauchschwalbe}}, \bibinfo {author} {\bibfnamefont {W.}~\bibnamefont {Lieke}}, \bibinfo {author} {\bibfnamefont {F.}~\bibnamefont {Steglich}}, \bibinfo {author} {\bibfnamefont {C.}~\bibnamefont {Godart}}, \bibinfo {author} {\bibfnamefont {L.}~\bibnamefont {Gupta}}, \ and\ \bibinfo {author} {\bibfnamefont {R.}~\bibnamefont {Parks}},\ }\href@noop {} {\bibfield  {journal} {\bibinfo  {journal} {Phys. Rev. B}\ }\textbf {\bibinfo {volume} {30}},\ \bibinfo {pages} {444} (\bibinfo {year} {1984})}\BibitemShut {NoStop}%
\bibitem [{\citenamefont {Yomo}\ \emph {et~al.}(1986)\citenamefont {Yomo}, \citenamefont {Hor}, \citenamefont {Meng},\ and\ \citenamefont {Chu}}]{yomo1986high}%
  \BibitemOpen
  \bibfield  {author} {\bibinfo {author} {\bibfnamefont {S.}~\bibnamefont {Yomo}}, \bibinfo {author} {\bibfnamefont {P.}~\bibnamefont {Hor}}, \bibinfo {author} {\bibfnamefont {R.}~\bibnamefont {Meng}}, \ and\ \bibinfo {author} {\bibfnamefont {C.}~\bibnamefont {Chu}},\ }\href@noop {} {\bibfield  {journal} {\bibinfo  {journal} {Journal of Magnetism and Magnetic Materials}\ }\textbf {\bibinfo {volume} {54}},\ \bibinfo {pages} {477} (\bibinfo {year} {1986})}\BibitemShut {NoStop}%
\bibitem [{\citenamefont {Gul{\'a}csi}\ and\ \citenamefont {Gul{\'a}csi}(1994)}]{gulacsi1994bcs}%
  \BibitemOpen
  \bibfield  {author} {\bibinfo {author} {\bibfnamefont {M.}~\bibnamefont {Gul{\'a}csi}}\ and\ \bibinfo {author} {\bibfnamefont {Z.}~\bibnamefont {Gul{\'a}csi}},\ }\href@noop {} {\bibfield  {journal} {\bibinfo  {journal} {Solid state communications}\ }\textbf {\bibinfo {volume} {90}},\ \bibinfo {pages} {51} (\bibinfo {year} {1994})}\BibitemShut {NoStop}%
\bibitem [{\citenamefont {Kishimoto}\ \emph {et~al.}(2003)\citenamefont {Kishimoto}, \citenamefont {Kawasaki},\ and\ \citenamefont {Ohno}}]{kishimoto2003mixed}%
  \BibitemOpen
  \bibfield  {author} {\bibinfo {author} {\bibfnamefont {Y.}~\bibnamefont {Kishimoto}}, \bibinfo {author} {\bibfnamefont {Y.}~\bibnamefont {Kawasaki}}, \ and\ \bibinfo {author} {\bibfnamefont {T.}~\bibnamefont {Ohno}},\ }\href@noop {} {\bibfield  {journal} {\bibinfo  {journal} {Physics letters A}\ }\textbf {\bibinfo {volume} {317}},\ \bibinfo {pages} {308} (\bibinfo {year} {2003})}\BibitemShut {NoStop}%
\bibitem [{\citenamefont {Ivashko}\ \emph {et~al.}(2025)\citenamefont {Ivashko}, \citenamefont {Igoa~Salda{\~n}a}, \citenamefont {Liu}, \citenamefont {Glaevecke}, \citenamefont {Gutowski}, \citenamefont {Nowak}, \citenamefont {K{\"o}hler}, \citenamefont {Winkler}, \citenamefont {Sch{\"o}ps}, \citenamefont {Schulte-Schrepping} \emph {et~al.}}]{ivashko2025p21}%
  \BibitemOpen
  \bibfield  {author} {\bibinfo {author} {\bibfnamefont {O.}~\bibnamefont {Ivashko}}, \bibinfo {author} {\bibfnamefont {F.}~\bibnamefont {Igoa~Salda{\~n}a}}, \bibinfo {author} {\bibfnamefont {J.}~\bibnamefont {Liu}}, \bibinfo {author} {\bibfnamefont {P.}~\bibnamefont {Glaevecke}}, \bibinfo {author} {\bibfnamefont {O.}~\bibnamefont {Gutowski}}, \bibinfo {author} {\bibfnamefont {R.}~\bibnamefont {Nowak}}, \bibinfo {author} {\bibfnamefont {K.}~\bibnamefont {K{\"o}hler}}, \bibinfo {author} {\bibfnamefont {B.}~\bibnamefont {Winkler}}, \bibinfo {author} {\bibfnamefont {A.}~\bibnamefont {Sch{\"o}ps}}, \bibinfo {author} {\bibfnamefont {H.}~\bibnamefont {Schulte-Schrepping}},  \emph {et~al.},\ }\href@noop {} {\bibfield  {journal} {\bibinfo  {journal} {Synchrotron Radiation}\ }\textbf {\bibinfo {volume} {32}} (\bibinfo {year} {2025})}\BibitemShut {NoStop}%
\bibitem [{\citenamefont {Misawa}\ \emph {et~al.}(2025)\citenamefont {Misawa}, \citenamefont {Kitou}, \citenamefont {Yamada}, \citenamefont {Gaggl}, \citenamefont {Nakano}, \citenamefont {Shibata}, \citenamefont {Okamura}, \citenamefont {Kriener}, \citenamefont {Baral}, \citenamefont {Nakamura} \emph {et~al.}}]{misawa2025successive}%
  \BibitemOpen
  \bibfield  {author} {\bibinfo {author} {\bibfnamefont {R.}~\bibnamefont {Misawa}}, \bibinfo {author} {\bibfnamefont {S.}~\bibnamefont {Kitou}}, \bibinfo {author} {\bibfnamefont {R.}~\bibnamefont {Yamada}}, \bibinfo {author} {\bibfnamefont {T.}~\bibnamefont {Gaggl}}, \bibinfo {author} {\bibfnamefont {R.}~\bibnamefont {Nakano}}, \bibinfo {author} {\bibfnamefont {Y.}~\bibnamefont {Shibata}}, \bibinfo {author} {\bibfnamefont {Y.}~\bibnamefont {Okamura}}, \bibinfo {author} {\bibfnamefont {M.}~\bibnamefont {Kriener}}, \bibinfo {author} {\bibfnamefont {P.~R.}\ \bibnamefont {Baral}}, \bibinfo {author} {\bibfnamefont {Y.}~\bibnamefont {Nakamura}},  \emph {et~al.},\ }\href@noop {} {\bibfield  {journal} {\bibinfo  {journal} {Adv. Mater.}\ ,\ \bibinfo {pages} {e13015}} (\bibinfo {year} {2025})}\BibitemShut {NoStop}%
\bibitem [{\citenamefont {Liu}\ \emph {et~al.}(2025)\citenamefont {Liu}, \citenamefont {Roppongi}, \citenamefont {Kimata}, \citenamefont {Ishihara}, \citenamefont {Grasset}, \citenamefont {Konczykowski}, \citenamefont {Ortiz}, \citenamefont {Wilson}, \citenamefont {Yoshimi}, \citenamefont {Shibauchi},\ and\ \citenamefont {Hashimoto}}]{liu2025}%
  \BibitemOpen
  \bibfield  {author} {\bibinfo {author} {\bibfnamefont {S.}~\bibnamefont {Liu}}, \bibinfo {author} {\bibfnamefont {M.}~\bibnamefont {Roppongi}}, \bibinfo {author} {\bibfnamefont {M.}~\bibnamefont {Kimata}}, \bibinfo {author} {\bibfnamefont {K.}~\bibnamefont {Ishihara}}, \bibinfo {author} {\bibfnamefont {R.}~\bibnamefont {Grasset}}, \bibinfo {author} {\bibfnamefont {M.}~\bibnamefont {Konczykowski}}, \bibinfo {author} {\bibfnamefont {B.~R.}\ \bibnamefont {Ortiz}}, \bibinfo {author} {\bibfnamefont {S.~D.}\ \bibnamefont {Wilson}}, \bibinfo {author} {\bibfnamefont {K.}~\bibnamefont {Yoshimi}}, \bibinfo {author} {\bibfnamefont {T.}~\bibnamefont {Shibauchi}}, \ and\ \bibinfo {author} {\bibfnamefont {K.}~\bibnamefont {Hashimoto}},\ }\href {\doibase 10.1103/d4dw-2v6k} {\bibfield  {journal} {\bibinfo  {journal} {Phys. Rev. Lett.}\ }\textbf {\bibinfo {volume} {135}},\ \bibinfo {pages} {056502} (\bibinfo {year} {2025})}\BibitemShut {NoStop}%
\bibitem [{\citenamefont {Xu}\ \emph {et~al.}(2025)\citenamefont {Xu}, \citenamefont {Le},\ and\ \citenamefont {Lin}}]{xu2025av3sb5}%
  \BibitemOpen
  \bibfield  {author} {\bibinfo {author} {\bibfnamefont {Z.}~\bibnamefont {Xu}}, \bibinfo {author} {\bibfnamefont {T.}~\bibnamefont {Le}}, \ and\ \bibinfo {author} {\bibfnamefont {X.}~\bibnamefont {Lin}},\ }\href@noop {} {\bibfield  {journal} {\bibinfo  {journal} {Chinese Physics Letters}\ } (\bibinfo {year} {2025})}\BibitemShut {NoStop}%
\bibitem [{\citenamefont {Mansart}\ \emph {et~al.}(2012)\citenamefont {Mansart}, \citenamefont {Cottet}, \citenamefont {Penfold}, \citenamefont {Dugdale}, \citenamefont {Tediosi}, \citenamefont {Chergui},\ and\ \citenamefont {Carbone}}]{mansart2012evidence}%
  \BibitemOpen
  \bibfield  {author} {\bibinfo {author} {\bibfnamefont {B.}~\bibnamefont {Mansart}}, \bibinfo {author} {\bibfnamefont {M.~J.}\ \bibnamefont {Cottet}}, \bibinfo {author} {\bibfnamefont {T.~J.}\ \bibnamefont {Penfold}}, \bibinfo {author} {\bibfnamefont {S.~B.}\ \bibnamefont {Dugdale}}, \bibinfo {author} {\bibfnamefont {R.}~\bibnamefont {Tediosi}}, \bibinfo {author} {\bibfnamefont {M.}~\bibnamefont {Chergui}}, \ and\ \bibinfo {author} {\bibfnamefont {F.}~\bibnamefont {Carbone}},\ }\href@noop {} {\bibfield  {journal} {\bibinfo  {journal} {Proceedings of the National Academy of Sciences}\ }\textbf {\bibinfo {volume} {109}},\ \bibinfo {pages} {5603} (\bibinfo {year} {2012})}\BibitemShut {NoStop}%
\bibitem [{\citenamefont {Ong}\ and\ \citenamefont {Monceau}(1977)}]{ong1977anomalous}%
  \BibitemOpen
  \bibfield  {author} {\bibinfo {author} {\bibfnamefont {N.}~\bibnamefont {Ong}}\ and\ \bibinfo {author} {\bibfnamefont {P.}~\bibnamefont {Monceau}},\ }\href@noop {} {\bibfield  {journal} {\bibinfo  {journal} {Phys. Rev. B}\ }\textbf {\bibinfo {volume} {16}},\ \bibinfo {pages} {3443} (\bibinfo {year} {1977})}\BibitemShut {NoStop}%
\bibitem [{\citenamefont {Ali}\ \emph {et~al.}(2014)\citenamefont {Ali}, \citenamefont {Xiong}, \citenamefont {Flynn}, \citenamefont {Tao}, \citenamefont {Gibson}, \citenamefont {Schoop}, \citenamefont {Liang}, \citenamefont {Haldolaarachchige}, \citenamefont {Hirschberger}, \citenamefont {Ong} \emph {et~al.}}]{ali2014large}%
  \BibitemOpen
  \bibfield  {author} {\bibinfo {author} {\bibfnamefont {M.~N.}\ \bibnamefont {Ali}}, \bibinfo {author} {\bibfnamefont {J.}~\bibnamefont {Xiong}}, \bibinfo {author} {\bibfnamefont {S.}~\bibnamefont {Flynn}}, \bibinfo {author} {\bibfnamefont {J.}~\bibnamefont {Tao}}, \bibinfo {author} {\bibfnamefont {Q.~D.}\ \bibnamefont {Gibson}}, \bibinfo {author} {\bibfnamefont {L.~M.}\ \bibnamefont {Schoop}}, \bibinfo {author} {\bibfnamefont {T.}~\bibnamefont {Liang}}, \bibinfo {author} {\bibfnamefont {N.}~\bibnamefont {Haldolaarachchige}}, \bibinfo {author} {\bibfnamefont {M.}~\bibnamefont {Hirschberger}}, \bibinfo {author} {\bibfnamefont {N.~P.}\ \bibnamefont {Ong}},  \emph {et~al.},\ }\href@noop {} {\bibfield  {journal} {\bibinfo  {journal} {Nature}\ }\textbf {\bibinfo {volume} {514}},\ \bibinfo {pages} {205} (\bibinfo {year} {2014})}\BibitemShut {NoStop}%
\bibitem [{\citenamefont {Qiu}\ \emph {et~al.}(2021)\citenamefont {Qiu}, \citenamefont {Gong}, \citenamefont {Wang}, \citenamefont {Zhang}, \citenamefont {Yang}, \citenamefont {Wang},\ and\ \citenamefont {Xiong}}]{qiu2021recent}%
  \BibitemOpen
  \bibfield  {author} {\bibinfo {author} {\bibfnamefont {D.}~\bibnamefont {Qiu}}, \bibinfo {author} {\bibfnamefont {C.}~\bibnamefont {Gong}}, \bibinfo {author} {\bibfnamefont {S.}~\bibnamefont {Wang}}, \bibinfo {author} {\bibfnamefont {M.}~\bibnamefont {Zhang}}, \bibinfo {author} {\bibfnamefont {C.}~\bibnamefont {Yang}}, \bibinfo {author} {\bibfnamefont {X.}~\bibnamefont {Wang}}, \ and\ \bibinfo {author} {\bibfnamefont {J.}~\bibnamefont {Xiong}},\ }\href@noop {} {\bibfield  {journal} {\bibinfo  {journal} {Adv. Mater.}\ }\textbf {\bibinfo {volume} {33}},\ \bibinfo {pages} {2006124} (\bibinfo {year} {2021})}\BibitemShut {NoStop}%
\bibitem [{\citenamefont {Sedlak}\ \emph {et~al.}(2012)\citenamefont {Sedlak}, \citenamefont {Scheuermann}, \citenamefont {Shiroka}, \citenamefont {Stoykov}, \citenamefont {Raselli},\ and\ \citenamefont {Amato}}]{sedlak2012musrsim}%
  \BibitemOpen
  \bibfield  {author} {\bibinfo {author} {\bibfnamefont {K.}~\bibnamefont {Sedlak}}, \bibinfo {author} {\bibfnamefont {R.}~\bibnamefont {Scheuermann}}, \bibinfo {author} {\bibfnamefont {T.}~\bibnamefont {Shiroka}}, \bibinfo {author} {\bibfnamefont {A.}~\bibnamefont {Stoykov}}, \bibinfo {author} {\bibfnamefont {A.}~\bibnamefont {Raselli}}, \ and\ \bibinfo {author} {\bibfnamefont {A.}~\bibnamefont {Amato}},\ }\href@noop {} {\bibfield  {journal} {\bibinfo  {journal} {Physics Procedia}\ }\textbf {\bibinfo {volume} {30}},\ \bibinfo {pages} {61} (\bibinfo {year} {2012})}\BibitemShut {NoStop}%
\bibitem [{\citenamefont {Brandt}(1988)}]{brandt1988flux}%
  \BibitemOpen
  \bibfield  {author} {\bibinfo {author} {\bibfnamefont {E.}~\bibnamefont {Brandt}},\ }\href@noop {} {\bibfield  {journal} {\bibinfo  {journal} {Phys. Rev. B}\ }\textbf {\bibinfo {volume} {37}},\ \bibinfo {pages} {2349} (\bibinfo {year} {1988})}\BibitemShut {NoStop}%
\bibitem [{\citenamefont {Taillefer}(2009)}]{taillefer2009fermi}%
  \BibitemOpen
  \bibfield  {author} {\bibinfo {author} {\bibfnamefont {L.}~\bibnamefont {Taillefer}},\ }\href@noop {} {\bibfield  {journal} {\bibinfo  {journal} {Journal of Physics: Condensed Matter}\ }\textbf {\bibinfo {volume} {21}},\ \bibinfo {pages} {164212} (\bibinfo {year} {2009})}\BibitemShut {NoStop}%
\bibitem [{\citenamefont {Xu}\ \emph {et~al.}(2021)\citenamefont {Xu}, \citenamefont {Yan}, \citenamefont {Yin}, \citenamefont {Xia}, \citenamefont {Fang}, \citenamefont {Chen}, \citenamefont {Li}, \citenamefont {Yang}, \citenamefont {Guo},\ and\ \citenamefont {Feng}}]{xu2021multiband}%
  \BibitemOpen
  \bibfield  {author} {\bibinfo {author} {\bibfnamefont {H.-S.}\ \bibnamefont {Xu}}, \bibinfo {author} {\bibfnamefont {Y.-J.}\ \bibnamefont {Yan}}, \bibinfo {author} {\bibfnamefont {R.}~\bibnamefont {Yin}}, \bibinfo {author} {\bibfnamefont {W.}~\bibnamefont {Xia}}, \bibinfo {author} {\bibfnamefont {S.}~\bibnamefont {Fang}}, \bibinfo {author} {\bibfnamefont {Z.}~\bibnamefont {Chen}}, \bibinfo {author} {\bibfnamefont {Y.}~\bibnamefont {Li}}, \bibinfo {author} {\bibfnamefont {W.}~\bibnamefont {Yang}}, \bibinfo {author} {\bibfnamefont {Y.}~\bibnamefont {Guo}}, \ and\ \bibinfo {author} {\bibfnamefont {D.-L.}\ \bibnamefont {Feng}},\ }\href@noop {} {\bibfield  {journal} {\bibinfo  {journal} {Phys. Rev. Lett.}\ }\textbf {\bibinfo {volume} {127}},\ \bibinfo {pages} {187004} (\bibinfo {year} {2021})}\BibitemShut {NoStop}%
\bibitem [{\citenamefont {Huang}\ and\ \citenamefont {Lin}(2020)}]{huang2020pairing}%
  \BibitemOpen
  \bibfield  {author} {\bibinfo {author} {\bibfnamefont {W.-M.}\ \bibnamefont {Huang}}\ and\ \bibinfo {author} {\bibfnamefont {H.-H.}\ \bibnamefont {Lin}},\ }\href@noop {} {\bibfield  {journal} {\bibinfo  {journal} {Scientific Reports}\ }\textbf {\bibinfo {volume} {10}},\ \bibinfo {pages} {7439} (\bibinfo {year} {2020})}\BibitemShut {NoStop}%
\bibitem [{\citenamefont {Eremin}\ and\ \citenamefont {Annett}(2006)}]{eremin2006magnetic}%
  \BibitemOpen
  \bibfield  {author} {\bibinfo {author} {\bibfnamefont {I.}~\bibnamefont {Eremin}}\ and\ \bibinfo {author} {\bibfnamefont {J.}~\bibnamefont {Annett}},\ }\href@noop {} {\bibfield  {journal} {\bibinfo  {journal} {Phys. Rev. B}\ }\textbf {\bibinfo {volume} {74}},\ \bibinfo {pages} {184524} (\bibinfo {year} {2006})}\BibitemShut {NoStop}%
\bibitem [{\citenamefont {Kresse}\ and\ \citenamefont {Furthm{\"u}ller}(1996)}]{VASP}%
  \BibitemOpen
  \bibfield  {author} {\bibinfo {author} {\bibfnamefont {G.}~\bibnamefont {Kresse}}\ and\ \bibinfo {author} {\bibfnamefont {J.}~\bibnamefont {Furthm{\"u}ller}},\ }\href {\doibase 10.1103/PhysRevB.54.11169} {\bibfield  {journal} {\bibinfo  {journal} {Phys. Rev. B}\ }\textbf {\bibinfo {volume} {54}},\ \bibinfo {pages} {11169} (\bibinfo {year} {1996})}\BibitemShut {NoStop}%
\bibitem [{\citenamefont {Bl\"ochl}(1994)}]{PAW}%
  \BibitemOpen
  \bibfield  {author} {\bibinfo {author} {\bibfnamefont {P.~E.}\ \bibnamefont {Bl\"ochl}},\ }\href {\doibase 10.1103/PhysRevB.50.17953} {\bibfield  {journal} {\bibinfo  {journal} {Phys. Rev. B}\ }\textbf {\bibinfo {volume} {50}},\ \bibinfo {pages} {17953} (\bibinfo {year} {1994})}\BibitemShut {NoStop}%
\bibitem [{\citenamefont {Perdew}\ \emph {et~al.}(2008)\citenamefont {Perdew}, \citenamefont {Ruzsinszky}, \citenamefont {Csonka}, \citenamefont {Vydrov}, \citenamefont {Scuseria}, \citenamefont {Constantin}, \citenamefont {Zhou},\ and\ \citenamefont {Burke}}]{PBEsol}%
  \BibitemOpen
  \bibfield  {author} {\bibinfo {author} {\bibfnamefont {J.~P.}\ \bibnamefont {Perdew}}, \bibinfo {author} {\bibfnamefont {A.}~\bibnamefont {Ruzsinszky}}, \bibinfo {author} {\bibfnamefont {G.~I.}\ \bibnamefont {Csonka}}, \bibinfo {author} {\bibfnamefont {O.~A.}\ \bibnamefont {Vydrov}}, \bibinfo {author} {\bibfnamefont {G.~E.}\ \bibnamefont {Scuseria}}, \bibinfo {author} {\bibfnamefont {L.~A.}\ \bibnamefont {Constantin}}, \bibinfo {author} {\bibfnamefont {X.}~\bibnamefont {Zhou}}, \ and\ \bibinfo {author} {\bibfnamefont {K.}~\bibnamefont {Burke}},\ }\href {\doibase 10.1103/PhysRevLett.100.136406} {\bibfield  {journal} {\bibinfo  {journal} {Phys. Rev. Lett.}\ }\textbf {\bibinfo {volume} {100}},\ \bibinfo {pages} {136406} (\bibinfo {year} {2008})}\BibitemShut {NoStop}%
\bibitem [{\citenamefont {Dudarev}\ \emph {et~al.}(1998)\citenamefont {Dudarev}, \citenamefont {Botton}, \citenamefont {Savrasov}, \citenamefont {Humphreys},\ and\ \citenamefont {Sutton}}]{DFT+U}%
  \BibitemOpen
  \bibfield  {author} {\bibinfo {author} {\bibfnamefont {S.~L.}\ \bibnamefont {Dudarev}}, \bibinfo {author} {\bibfnamefont {G.~A.}\ \bibnamefont {Botton}}, \bibinfo {author} {\bibfnamefont {S.~Y.}\ \bibnamefont {Savrasov}}, \bibinfo {author} {\bibfnamefont {C.~J.}\ \bibnamefont {Humphreys}}, \ and\ \bibinfo {author} {\bibfnamefont {A.~P.}\ \bibnamefont {Sutton}},\ }\href {\doibase 10.1103/PhysRevB.57.1505} {\bibfield  {journal} {\bibinfo  {journal} {Phys. Rev. B}\ }\textbf {\bibinfo {volume} {57}},\ \bibinfo {pages} {1505} (\bibinfo {year} {1998})}\BibitemShut {NoStop}%
\bibitem [{\citenamefont {Lloyd-Williams}\ and\ \citenamefont {Monserrat}(2015)}]{nondiagonal_supercells}%
  \BibitemOpen
  \bibfield  {author} {\bibinfo {author} {\bibfnamefont {J.~H.}\ \bibnamefont {Lloyd-Williams}}\ and\ \bibinfo {author} {\bibfnamefont {B.}~\bibnamefont {Monserrat}},\ }\href {\doibase 10.1103/PhysRevB.92.184301} {\bibfield  {journal} {\bibinfo  {journal} {Phys. Rev. B}\ }\textbf {\bibinfo {volume} {92}},\ \bibinfo {pages} {184301} (\bibinfo {year} {2015})}\BibitemShut {NoStop}%
\bibitem [{\citenamefont {Amato}\ \emph {et~al.}(2017)\citenamefont {Amato}, \citenamefont {Luetkens}, \citenamefont {Sedlak}, \citenamefont {Stoykov}, \citenamefont {Scheuermann}, \citenamefont {Elender}, \citenamefont {Raselli},\ and\ \citenamefont {Graf}}]{amato2017new}%
  \BibitemOpen
  \bibfield  {author} {\bibinfo {author} {\bibfnamefont {A.}~\bibnamefont {Amato}}, \bibinfo {author} {\bibfnamefont {H.}~\bibnamefont {Luetkens}}, \bibinfo {author} {\bibfnamefont {K.}~\bibnamefont {Sedlak}}, \bibinfo {author} {\bibfnamefont {A.}~\bibnamefont {Stoykov}}, \bibinfo {author} {\bibfnamefont {R.}~\bibnamefont {Scheuermann}}, \bibinfo {author} {\bibfnamefont {M.}~\bibnamefont {Elender}}, \bibinfo {author} {\bibfnamefont {A.}~\bibnamefont {Raselli}}, \ and\ \bibinfo {author} {\bibfnamefont {D.}~\bibnamefont {Graf}},\ }\href {\doibase 10.1063/1.4986045} {\bibfield  {journal} {\bibinfo  {journal} {Review of Scientific Instruments}\ }\textbf {\bibinfo {volume} {88}} (\bibinfo {year} {2017}),\ 10.1063/1.4986045}\BibitemShut {NoStop}%
\bibitem [{\citenamefont {Khasanov}\ \emph {et~al.}(2005)\citenamefont {Khasanov}, \citenamefont {Eshchenko}, \citenamefont {Di~Castro}, \citenamefont {Shengelaya}, \citenamefont {La~Mattina}, \citenamefont {Maisuradze}, \citenamefont {Baines}, \citenamefont {Luetkens}, \citenamefont {Karpinski}, \citenamefont {Kazakov},\ and\ \citenamefont {Keller}}]{Khasanov104504}%
  \BibitemOpen
  \bibfield  {author} {\bibinfo {author} {\bibfnamefont {R.}~\bibnamefont {Khasanov}}, \bibinfo {author} {\bibfnamefont {D.~G.}\ \bibnamefont {Eshchenko}}, \bibinfo {author} {\bibfnamefont {D.}~\bibnamefont {Di~Castro}}, \bibinfo {author} {\bibfnamefont {A.}~\bibnamefont {Shengelaya}}, \bibinfo {author} {\bibfnamefont {F.}~\bibnamefont {La~Mattina}}, \bibinfo {author} {\bibfnamefont {A.}~\bibnamefont {Maisuradze}}, \bibinfo {author} {\bibfnamefont {C.}~\bibnamefont {Baines}}, \bibinfo {author} {\bibfnamefont {H.}~\bibnamefont {Luetkens}}, \bibinfo {author} {\bibfnamefont {J.}~\bibnamefont {Karpinski}}, \bibinfo {author} {\bibfnamefont {S.~M.}\ \bibnamefont {Kazakov}}, \ and\ \bibinfo {author} {\bibfnamefont {H.}~\bibnamefont {Keller}},\ }\href {\doibase 10.1103/PhysRevB.72.104504} {\bibfield  {journal} {\bibinfo  {journal} {Phys. Rev. B}\ }\textbf {\bibinfo {volume} {72}},\ \bibinfo {pages} {104504} (\bibinfo {year} {2005})}\BibitemShut {NoStop}%
\bibitem [{\citenamefont {Suter}\ and\ \citenamefont {Wojek}(2012)}]{Suter69}%
  \BibitemOpen
  \bibfield  {author} {\bibinfo {author} {\bibfnamefont {A.}~\bibnamefont {Suter}}\ and\ \bibinfo {author} {\bibfnamefont {B.}~\bibnamefont {Wojek}},\ }\href {\doibase https://doi.org/10.1016/j.phpro.2012.04.042} {\bibfield  {journal} {\bibinfo  {journal} {Phys. Procedia}\ }\textbf {\bibinfo {volume} {30}},\ \bibinfo {pages} {69} (\bibinfo {year} {2012})}\BibitemShut {NoStop}%
\bibitem [{\citenamefont {Tinkham}(2004)}]{Tinkham2004}%
  \BibitemOpen
  \bibfield  {author} {\bibinfo {author} {\bibfnamefont {M.}~\bibnamefont {Tinkham}},\ }\href {http://app.knovel.com/hotlink/toc/id:kpISE00023/introduction-to-superconductivity} {\emph {\bibinfo {title} {Introduction to superconductivity}}},\ \bibinfo {edition} {2nd}\ ed.\ (\bibinfo  {publisher} {Dover Publications Mineola, N.Y},\ \bibinfo {address} {Mineola, N.Y},\ \bibinfo {year} {2004})\BibitemShut {NoStop}%
\bibitem [{\citenamefont {Carrington}\ and\ \citenamefont {Manzano}(2003)}]{Carrington205}%
  \BibitemOpen
  \bibfield  {author} {\bibinfo {author} {\bibfnamefont {A.}~\bibnamefont {Carrington}}\ and\ \bibinfo {author} {\bibfnamefont {F.}~\bibnamefont {Manzano}},\ }\href {\doibase https://doi.org/10.1016/S0921-4534(02)02319-5} {\bibfield  {journal} {\bibinfo  {journal} {Physica C: Superconductivity}\ }\textbf {\bibinfo {volume} {385}},\ \bibinfo {pages} {205} (\bibinfo {year} {2003})}\BibitemShut {NoStop}%
\bibitem [{\citenamefont {Guguchia}\ \emph {et~al.}(2015)\citenamefont {Guguchia}, \citenamefont {Amato}, \citenamefont {Kang}, \citenamefont {Luetkens}, \citenamefont {Biswas}, \citenamefont {Prando}, \citenamefont {von Rohr}, \citenamefont {Bukowski}, \citenamefont {Shengelaya}, \citenamefont {Keller} \emph {et~al.}}]{guguchia2015direct}%
  \BibitemOpen
  \bibfield  {author} {\bibinfo {author} {\bibfnamefont {Z.}~\bibnamefont {Guguchia}}, \bibinfo {author} {\bibfnamefont {A.}~\bibnamefont {Amato}}, \bibinfo {author} {\bibfnamefont {J.}~\bibnamefont {Kang}}, \bibinfo {author} {\bibfnamefont {H.}~\bibnamefont {Luetkens}}, \bibinfo {author} {\bibfnamefont {P.~K.}\ \bibnamefont {Biswas}}, \bibinfo {author} {\bibfnamefont {G.}~\bibnamefont {Prando}}, \bibinfo {author} {\bibfnamefont {F.}~\bibnamefont {von Rohr}}, \bibinfo {author} {\bibfnamefont {Z.}~\bibnamefont {Bukowski}}, \bibinfo {author} {\bibfnamefont {A.}~\bibnamefont {Shengelaya}}, \bibinfo {author} {\bibfnamefont {H.}~\bibnamefont {Keller}},  \emph {et~al.},\ }\href@noop {} {\bibfield  {journal} {\bibinfo  {journal} {Nat. Commun.}\ }\textbf {\bibinfo {volume} {6}},\ \bibinfo {pages} {8863} (\bibinfo {year} {2015})}\BibitemShut {NoStop}%
\end{thebibliography}%
\end{document}